# State-of-the-Art Perovskite Solar Cells Benefit from Photon Recycling at Maximum Power Point

Roberto Brenes[1,3,†], Madeleine Laitz[1,3,†], Joel Jean[1,2], Dane W. deQuilettes[3], Vladimir Bulović[1,3*]

[1]Department of Electrical Engineering and Computer Science, Massachusetts Institute of Technology, 77 Massachusetts Avenue, Cambridge, Massachusetts 02139, USA
[2]Swift Solar Inc., Golden, Colorado 80401, USA
[3]Research Laboratory of Electronics, Massachusetts Institute of Technology, 77 Massachusetts Avenue, Cambridge, Massachusetts 02139, USA

**†Equal Contribution**

*Corresponding Author: bulovic@mit.edu

**Abstract**

**Photon recycling is required for a solar cell to achieve an open-circuit voltage ($V_{OC}$) and power conversion efficiency (PCE) approaching the Shockley-Queisser theoretical limit. In metal halide perovskite solar cells, the achievable performance gains from photon recycling remain uncertain due to high variability in perovskite material quality and the non-radiative recombination rate ($k_1$). In this work, we study state-of-the-art $Cs_{0.05}(MA_{0.17}FA_{0.83})_{0.95}Pb(I_{0.83}Br_{0.17})_3$ films and analyze the impact of varying non-radiative recombination rates on photon recycling and device performance. Importantly, we predict the impact of photon recycling at the maximum power point (MPP), demonstrating an absolute PCE increase of up to 2.0% in the radiative limit, primarily due to a 77 mV increase in $V_{MPP}$. Even with finite non-radiative recombination, benefits from photon recycling can be achieved when non-radiative lifetimes and external LED electroluminescence efficiencies measured at open-circuit, $Q_e^{LED}(V_{OC})$, exceed 2 μs and 10%, respectively. This analysis clarifies the opportunity to fully exploit photon recycling to push the real-world performance of perovskite solar cells toward theoretical limits.**

Improving solar cell power conversion efficiency (PCE) requires both optimization of device architectures and an understanding of fundamental photophysics. For example, GaAs cell efficiencies increased rapidly from 25.1% to 28.8% through photon management and efficient light extraction [1–4]. In high-quality optoelectronic materials like GaAs, photons can undergo multiple absorption and emission events before escaping, a phenomenon called photon recycling. Recycling increases the charge carrier density at steady-state and results in a higher quasi-Fermi level splitting and voltage ($\mu = qV_{OC}$) [4–7]. It also slows external emission, decreasing the radiative saturation current [8,9]. Together, these effects can boost the performance of high-efficiency solar cells toward the Shockley-Queisser theoretical limit. To take advantage of photon recycling, a photovoltaic (PV) absorber material must exhibit a small Stokes shift, strong band-edge absorption, and, most importantly, high photoluminescence quantum efficiency (PLQE) [3–5,10–12].

Relatively low PLQEs have thus far limited the extent of photon recycling observed in perovskite thin films and single crystals [13–15]. Low PLQEs of <15% result from high first-order non-radiative recombination rates ($k_1 \sim 10^6 - 10^9$ s$^{-1}$) and trap state densities ($10^{15} - 10^{17}$ cm$^{-3}$) [16–18]. For example, Pazos-Outón *et al.* demonstrated that the average photon only undergoes one recycling event in a typical $CH_3NH_3PbI_3$ film, but predicted that up to 25 recycling events could be sustained with a sufficiently high-quality sample [13]. This is in contrast to GaAs used in state-of-the-art PV devices, where an average photon can participate in up to 50 recycling events [3]. In thick perovskite single crystals, several reports have used time-resolved photoluminescence spectroscopy to show a characteristic red-shift in emission spectra over time due to photon recycling [13]. This phenomenon is expected to be efficient in single crystals, which exhibit low bulk defect densities ($\sim 10^{10}$ cm$^{-3}$). However, the high surface recombination velocities of 5,800 cm s$^{-1}$ lead to rapid quenching of excess carriers, diminishing the probability of photon recycling [14,19–23].

Passivated perovskite thin films with record-low non-radiative recombination rates and defect densities have achieved internal PLQEs exceeding 90% [18,24,25] — approaching the highest-quality double-heterostructured GaAs films [26]. These recent advances in material quality theoretically enable improved photon recycling and management in perovskite devices, but thus far it has been unclear how to realize practical efficiency gains. Futhermore, other works have only considered the impact of photon recycling on perovskite solar cells at open circuit, where the extent of recycling differs significantly from operation at the maximum power point (MPP). Under operation, rapid charge extraction reduces the steady-state carrier density, allowing non-radiative processes to compete with radiative recombination and photon recycling. The practical importance of photon recycling in perovskite solar cells thus remains unclear.

Here we perform a theoretical analysis of the impact of photon recycling in state-of-the-art $Cs_{0.05}(MA_{0.17}FA_{0.83})_{0.95}Pb(I_{0.83}Br_{0.17})_3$ (triple-cation) films at maximum-power-point with varying first-order non-radiative recombination rates ($k_1$) and external emission efficiencies. [27,28]. Our model reveals the changes in carrier density and luminescence efficiency at maximum-power-point attributable to photon recycling and identifies optoelectronic material quality targets— i.e., external luminescence quantum efficiency and non-radiative recombination rates – toward which the community can strive. Quantifying these values is critical, as several reports have shown that devices with low non-radiative recombination can achieve $V_{OC}$ deficits below 0.4 V, which is the deficit regime in which GaAs began to benefit from photon recycling [29–33].

To quantify the effect of photon recycling on device performance, current-voltage (J-V) curves were simulated using a detailed balance model and experimentally determined absorption coefficient and refractive index data for $Cs_{0.05}(MA_{0.17}FA_{0.83})_{0.95}Pb(I_{0.83}Br_{0.17})_3$ (S2) [16,27]. In the Supporting Information (SI), we discuss key assumptions used in the model, which are consistent with previous analyses [7,27,34–37].

We first perform a detailed-balance calculation in the radiative limit (i.e., no non-radiative recombination) by equating the generation current with the recombination and extraction currents. The total current ($J_{total}$) as a function of voltage (V) is then defined as follows:

$$[1] \quad J_{total}(V) = J_{SC} - J_0^{rad,ext}(V)$$

where $J_{SC}$ is the short-circuit current density

$$[2] \quad J_{SC} = q \int_0^\infty a(E)\phi_{sun}(E)dE$$

and $J_0^{rad,ext}$ is the radiative saturation current

$$[3] \quad J_0^{rad,ext}(V) = q\pi e^{qV/kT} \int_0^\infty a(E)\phi_{bb}(E)dE$$

where $q$ is the fundamental charge, $k$ is the Boltzmann constant, $T$ is the cell temperature, $a(E)$ is the absorptivity, $\phi_{sun}(E)$ is the AM1.5 spectral photon flux, and $\phi_{bb}(E)$ is the blackbody spectral photon flux all as a function of energy.

Importantly, and as discussed in depth previously [36,37], photon recycling is implicit in the Shockley-Queisser detailed balance calculation, where the total photon flux emitted from the front surface of the device is used to determine the external radiative saturation current: $J_0^{rad,ext}$ (Eq. 3). Here, only the emitted photons in the escape cone contribute to $J_0^{rad,ext}$, regardless of the number of photon recycling events before escape [34,38].

Next, we determine the benefits of photon recycling by considering the radiative saturation current when photon recycling is not included in the calculation [34,39]. The internal radiative saturation current ($J_0^{rad,int}$) is similar to the external radiative saturation current but is enhanced by the photon mode density within the dielectric medium and integrated over the sphere of emission [36]. Equation 4 describes a microscopic view of recombination *within* the active region where all photons immediately escape into the surrounding environment [34,38]:

$$[4] \quad J_0^{\text{rad,int}} = qe^{qV/kT} \int_0^\infty 4\pi n_r^2(E)\alpha(E)\,\phi_{bb}(E)\,dE$$

where $n_r(E)$ is the index of refraction and $\alpha(E)$ is the absorption coefficient. As a proof of concept, we use Equations 3 and 4 to calculate the internal and external second-order radiative recombination rate constants for a triple-cation film with an intrinsic carrier density of $n_i$ = 2.3×10$^5$ cm$^{-3}$: $k_2^{int}$ = 2×10$^{-10}$ cm$^3$ s$^{-1}$ and $k_2^{ext}$ = 1×10$^{-11}$ cm$^3$ s$^{-1}$, respectively, which agree well with experimental reports considering photon recycling effects [24,40–43].

In order to make our calculations relevant for real-world perovskite devices exhibiting disorder and band-tailing, we experimentally determined the energy-dependent absorption coefficient and refractive index using photothermal deflection spectroscopy (PDS) and ellipsometry. Figure 1 shows the theoretical current-voltage (J-V) curves for a triple-cation perovskite solar cell (see SI for CH$_3$NH$_3$PbI$_3$) in the radiative limit with and without photon recycling, calculated using the external and internal radiative saturation currents, respectively. We highlight that the calculated maximum efficiency with photon recycling (30.2%) corroborates previously reported theoretical limits for a CH$_3$NH$_3$PbI$_3$ perovskite solar cell of a similar bandgap [27], where the theoretical J$_{SC}$ (25.8 mA cm$^{-2}$) is only slightly higher than what has been achieved experimentally (24.5 mA cm$^{-2}$ [44]). These results emphasize the need to optimize V$_{oc}$ and fill-factor (FF) through reducing non-radiative recombination and harnessing photon recycling.

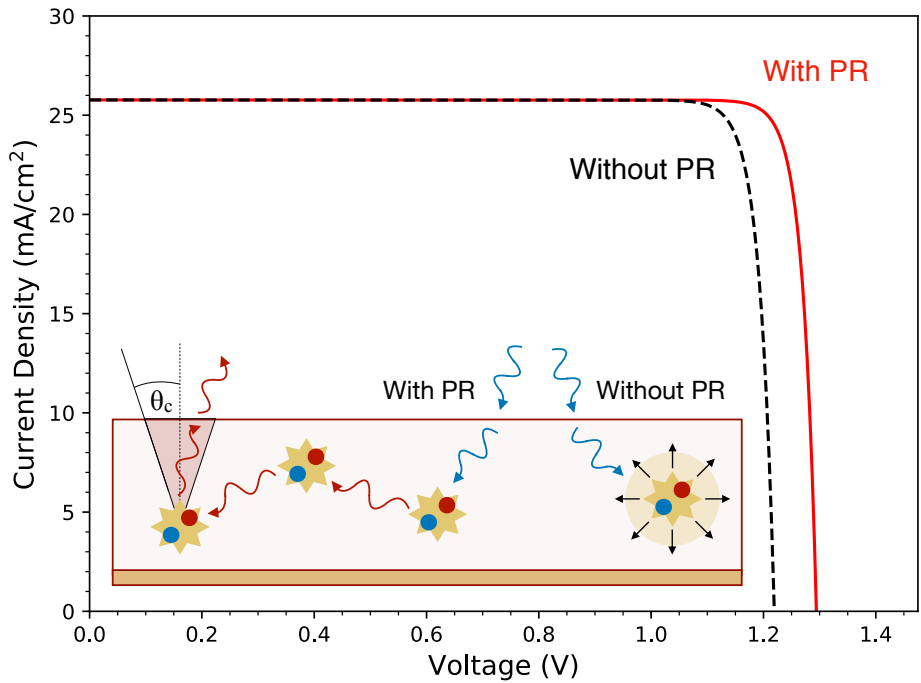

*Figure 1. Detailed-balance simulation of J-V curves for an ideal Cs$_{0.05}$(MA$_{0.17}$FA$_{0.83}$)$_{0.95}$Pb(I$_{0.83}$Br$_{0.17}$)$_3$ perovskite photovoltaic device in the radiative limit (no non-radiative recombination) with (red trace) and without (black dashed trace) photon recycling.*

|  | $J_{SC}$ [mA/cm$^2$] | $V_{OC}$ [V] | FF | $J_{MPP}$ [mA/cm$^2$] | $V_{MPP}$ [V] | PCE [%] |
|---|---|---|---|---|---|---|
| **No PR** | 25.77 | 1.22 | 0.899 | 25.19 | 1.12 | 28.2 |
| **With PR** | 25.77 | 1.29 | 0.906 | 25.23 | 1.20 | 30.2 |

*Table 1. J-V characteristics and power conversion efficiency (PCE) extracted from the simulated J-V curve in the radiative limit with and without photon recycling (PR). Photon recycling leads primarily to enhancements in operating voltage ($V_{MPP}$).*

Table 1 shows that photon recycling improves PV device performance at both open circuit and MPP. The $V_{OC}$ increase of $\Delta V_{OC}^{PR}$ = 70 mV calculated in this work is consistent with the 70 mV value predicted by Kirchartz *et al.* for a planar device architecture with Beer-Lambert absorption [27]. Extending beyond previous studies, our full J-V simulation also shows that photon recycling improves the maximum-power-point voltage $V_{MPP}$ ($\Delta V_{MPP}^{PR}$ = 80 mV) and the fill factor (FF), producing an absolute increase in PCE of 2.0%. We note that the short-circuit current density remains unchanged because, with or without photon recycling, $J_{SC}$ only depends on the absorptivity of the material and the solar irradiance.

One highlight of this analysis is that the maximum $V_{OC}$ achievable without photon recycling is only 1.22 V for both the triple cation and $CH_3NH_3PbI_3$ films (SI) — a voltage deficit of 0.38 V for each formulation. Our results suggest that any perovskite device (bandgap ~ 1.6 eV) with $V_{OC}$ > 1.22 V and $V_{MPP}$ > 1.12 V benefits from photon recycling [29,45]. In this regard, Liu *et al.* recently reported a record-setting open-circuit voltage of 1.26 V for $CH_3NH_3PbI_3$, where our calculations predict ~40 meV can be attributed to photon recycling effects alone.

Theoretical photovoltaic performance limits are useful for setting efficiency targets, but most absorber layers perform far from the radiative limit due to non-radiative losses. Perovskites are no exception — typical films exhibit PLQEs of <15% at 1-sun equivalent generation, with first-order non-radiative recombination rates ranging from $k_1$ ~ 10$^6$ – 10$^9$ s$^{-1}$, depending on chemical composition and processing methods [13,16,17]. However, with recently developed passivation techniques, $k_1$ values have been decreasing and will likely continue to decrease as passivation mechanisms are better understood and implemented [14]. For example, a low non-radiative recombination rate of $k_1$ = 1.7×10$^5$ s$^{-1}$ has been reported for tri-*n*-octylphosphine oxide (TOPO)-treated $CH_3NH_3PbI_3$ films [18].

To adapt our model to non-ideal scenarios, we modify the saturation current density (Eq. 3 and Eq. 4) to account for non-radiative Shockley-Read-Hall (SRH) and Auger recombination, as previously reported by Pazos-Outón *et al.* (Eq. 5) [27]:

$$[5] \quad J_0 = J_0^{rad} + J_0^{nonrad} = J_0^{rad} + J_{SRH} + J_A$$

where $J_0^{rad}$ is the radiative recombination rate (external or internal) and $J_0^{nonrad}$ is the sum of the non-radiative, first-order SRH ($J_{SRH}$) and non-radiative, third-order Auger ($J_A$) recombination currents. The SRH and Auger recombination rates are described for a carrier density ($n$) in quasi-thermal equilibrium using the law of mass action (Eq. 6):

$$[6] \quad n(V) = n_i e^{qV/2kT}$$

$$[7] \quad J_{SRH}(V) = qk_1 n(V) d$$

$$[8] \quad J_A(V) = qk_3 n(V)^3 d$$

where $n_i$ is the intrinsic carrier density, $d$ is the film thickness, and $k_1$ and $k_3$ are the first-order SRH and third-order Auger recombination rate constants, respectively.

Figure 2 shows the impact of different non-radiative recombination values on device performance with and without photon recycling. Figure 2a shows the simulated J-V curves with $k_1$ = 1×10$^4$ s$^{-1}$, which closely resembles those reported in Figure 1 ($k_1$ = 0), suggesting that radiative recombination outcompetes non-radiative recombination and photon recycling is almost completely exploited at this low $k_1$. Figure 2b shows that, as $k_1$ increases to 2×10$^5$ s$^{-1}$, the effect of photon recycling is greatly reduced and eventually becomes negligible when non-radiative rates approach 3×10$^6$ s$^{-1}$ (Figure 2c).

Figure 2d shows both the $V_{OC}$ and $V_{MPP}$ with and without photon recycling for varying $k_1$ values. For $k_1$ exceeding a threshold value of 2×10$^6$ s$^{-1}$ (i.e. $\tau_1$ < 500 ns), we observe no increase in $V_{OC}$ and $V_{MPP}$ with photon recycling. For $k_1$ between 7×10$^5$ s$^{-1}$ and 2×10$^6$ s$^{-1}$ (i.e. $\tau_1$ 500-1430 ns), photon recycling can improve $V_{OC}$ but the fill-factor decreases and, therefore, PCE enhancements are negligible (S3). With the full J-V simulation we can see that $V_{MPP}$ is unaffected at these values. Only when $k_1$ is reduced below 7×10$^5$ s$^{-1}$ ($\tau_1$ > 1430 ns) does photon recycling improve the MPP and efficiency (S3 and S4). For example, at $k_1$ = 2×10$^5$ s$^{-1}$ (Figure 2b), photon recycling increases $V_{OC}$ by 50 mV but $V_{MPP}$ by only 20 mV (Table 2). It is also interesting to note that for $k_1$= 3x10$^6$ s$^{-1}$, the PCE is comparable to current record-performing devices [46].

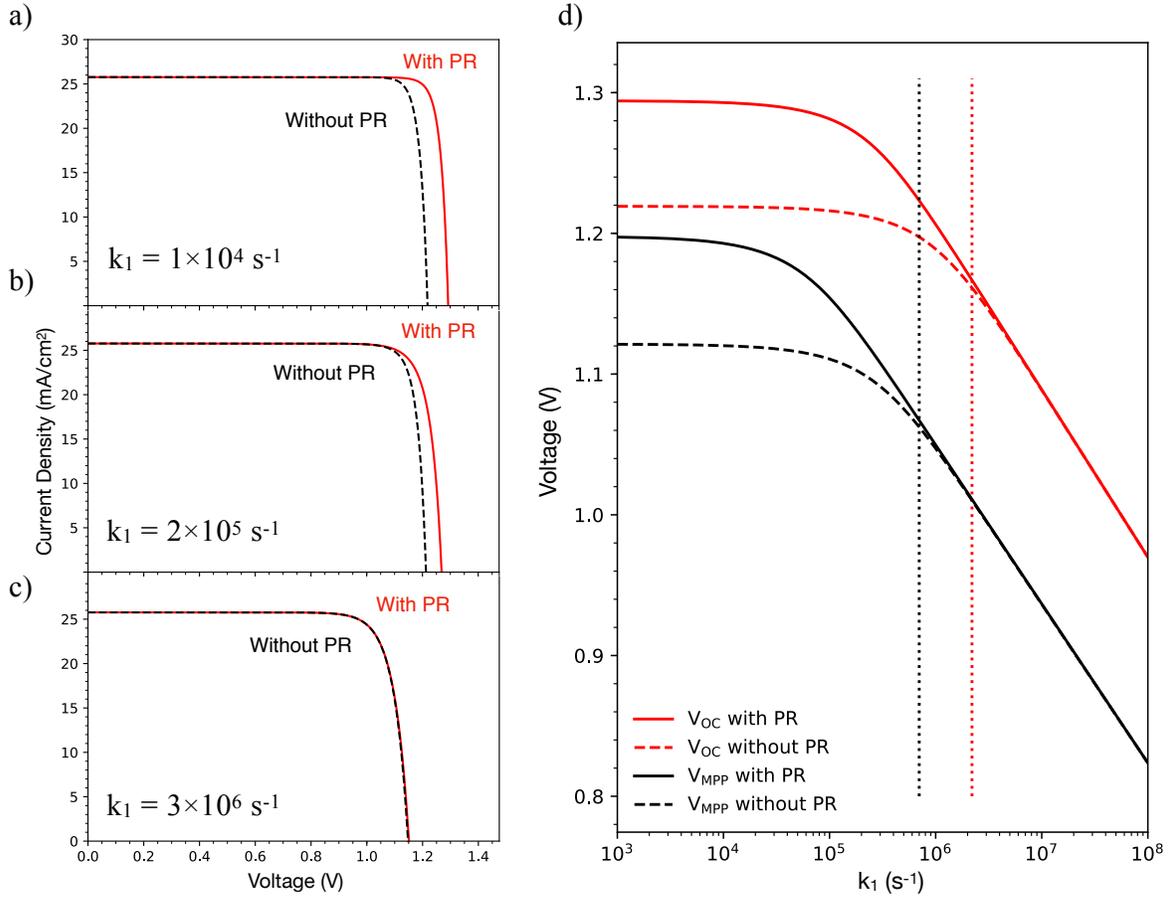

*Figure 2.* Simulated J-V curves (triple cation, 1.6 eV bandgap) with and without photon recycling (PR) for $k_1$ values of (a) $1\times10^4$ $s^{-1}$, (b) $2\times10^5$ $s^{-1}$, and (c) $3\times10^6$ $s^{-1}$ ($k_2^{int}$ = $2.0\times10^{-10}$ $cm^3$ $s^{-1}$ and $k_3$ = $1\times10^{-28}$ $cm^6$ $s^{-1}$). (d) $V_{OC}$ (red lines) and $V_{MPP}$ (black lines) as a function of $k_1$, revealing differences in the onset of performance improvements due to PR. Dotted vertical red and black lines indicate $k_1$ thresholds ($2\times10^6$ $s^{-1}$ and $7\times10^5$ $s^{-1}$, respectively) below which PR improves performance at open-circuit and MPP, respectively.

| | | $J_{SC}$ [mA/cm$^2$] | $V_{OC}$ [V] | FF | $J_{MPP}$ [mA/cm$^2$] | $V_{MPP}$ [V] | PCE [%] |
|---|---|---|---|---|---|---|---|
| $k_1$ = $1\times10^4$ $s^{-1}$ | No PR | 25.77 | 1.22 | 0.897 | 25.13 | 1.12 | 28.2 |
| | With PR | 25.77 | 1.29 | 0.900 | 25.17 | 1.19 | 30.0 |
| $k_1$ = $2\times10^5$ $s^{-1}$ | No PR | 25.77 | 1.21 | 0.876 | 24.86 | 1.10 | 27.4 |
| | With PR | 25.77 | 1.27 | 0.850 | 24.67 | 1.13 | 27.8 |
| $k_1$ = $3\times10^6$ $s^{-1}$ | No PR | 25.77 | 1.15 | 0.824 | 24.50 | 1.00 | 24.4 |
| | With PR | 25.77 | 1.15 | 0.822 | 24.50 | 1.00 | 24.4 |

*Table 2.* Device parameters extracted from the simulated J-V curves for $k_1$ = $1\times10^4$, $2\times10^5$, and $3\times10^6$ $s^{-1}$ with and without photon recycling (PR).

To better understand the recombination processes governing PV device behavior with and without photon recycling, we break down the J-V curve from Figure 2b ($k_1 = 2\times10^5$ s$^{-1}$) into its individual recombination components. The absolute magnitude is calculated using Equations 3-4 and 7-8, and the fraction of each recombination mechanism is its magnitude divided by the total recombination current (i.e. $J_{SRH,rad,A}/J_{tot}$).

Figures 3a and b show $J_0^{rad,int}$ and $J_0^{rad,ext}$ as a function of voltage, along with $J_{SRH}$ and $J_A$, which are non-radiative pathways and have the same functional form with or without photon recycling. We observe that photon recycling shifts radiative recombination to higher onset voltages and therefore reduces the magnitude of the radiative saturation current at MPP. Figures 3c and d give further insight into these results, where the fractions of each recombination current are compared as a function of voltage. With photon recycling, SRH recombination becomes the limiting pathway (solid blue and red traces).

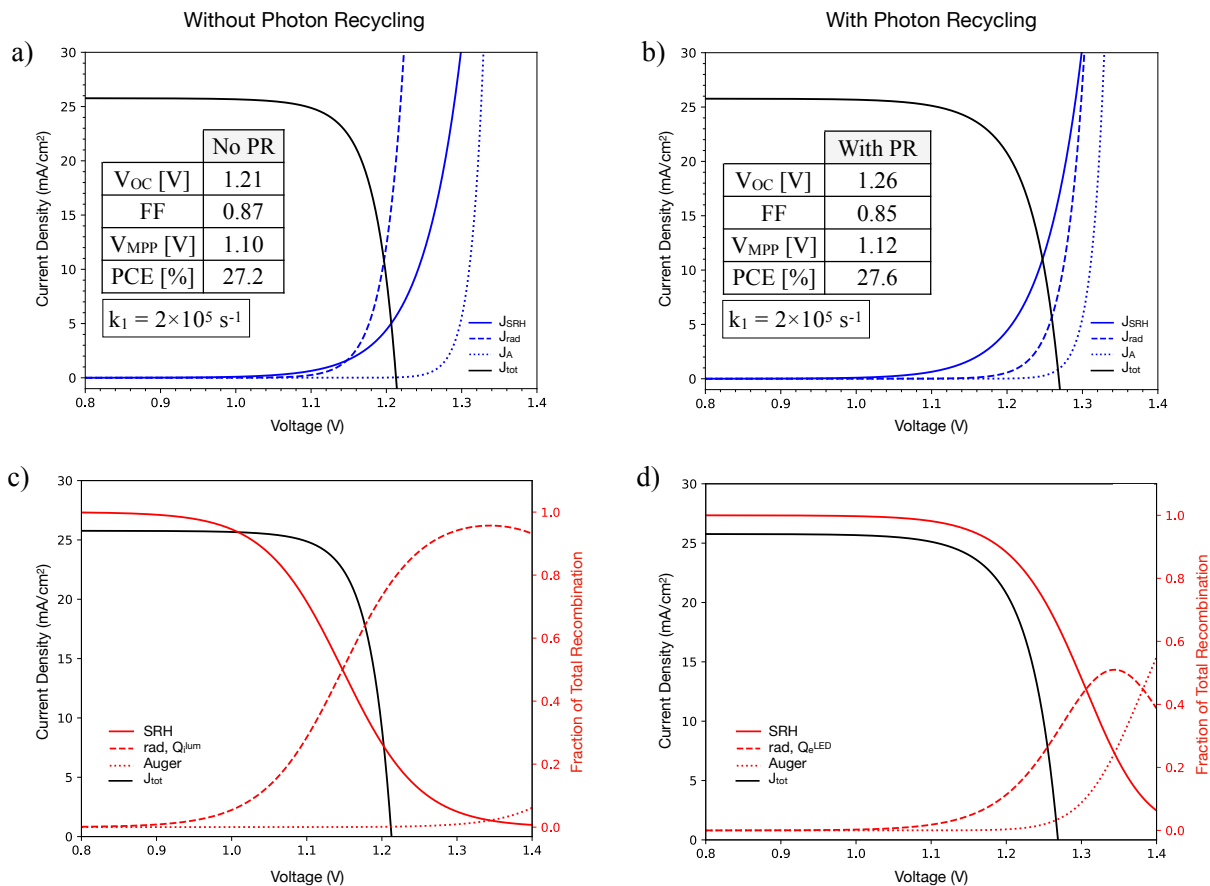

*Figure 3.* Simulated J-V curves (black traces) for $k_1 = 2\times10^5$ s$^{-1}$ (a, c) without and (b, d) with photon recycling (PR) are shown with the magnitude of SRH, radiative, and Auger recombination currents as a function of voltage (blue traces). (c,d) The fractions of total recombination current due to SRH, radiative, and Auger recombination are shown at each voltage (red traces) (c) without and (d) with PR. The fraction of radiative recombination as a function of voltage with and without PR is equivalent to $Q_e^{LED}$ and $Q_i^{lum}$, respectively.

We note that the radiative recombination fraction without photon recycling in Figure 3c is equal to the internal luminescence quantum efficiency ($Q_i^{lum}$), which has similarly been defined elsewhere [36]:

$$[9] \quad Q_i^{lum}(V) = \frac{J_0^{rad,int}(V)}{J_{SRH}(V) + J_0^{rad,int}(V) + J_A(V)}$$

The radiative recombination fraction with photon recycling in Figure 3d yields the external light-emitting diode (LED) electroluminescence efficiency ($Q_e^{LED}$), which is connected to the mean probability of photon escape from the film ($P_{esc}$) through a geometric series [36]:

$$[10] \quad Q_e^{LED}(V) = \frac{P_{esc} Q_i^{lum}(V)}{1 - Q_i^{lum}(V)(1 - P_{esc})}$$

Here, the escape probability ($P_{esc}$) can be defined as the ratio of the external to the internal radiative saturation current (Eq. 11) [36]:

$$[11] \quad P_{esc} = \frac{J_0^{rad,ext}}{J_0^{rad,int}}$$

$Q_e^{LED}$ is a function of the injection current, and thus it is necessary to denote both the injection current and corresponding voltage at which the current is achieved for a given $Q_e^{LED}$. Due to the reciprocity relations that link optical output to electrical input, $Q_e^{LED}$ values are often measured at an injection current equivalent to the photocurrent [47,48]. Unless otherwise stated, we report $Q_e^{LED}$ values calculated with an injection current equivalent to $J_{SC}$ – i.e. open-circuit voltage, $Q_e^{LED}(V_{OC})$. Considering Equation 10, Figure 3d shows that a device with $k_1$ = 2×10$^5$ s$^{-1}$ (i.e. $\tau_1$ = 5 µs) should demonstrate a $Q_e^{LED}(V_{OC})$ of 31.5%. Importantly, the external emission efficiency of a solar cell is a metric that has been shown to directly correlate with power conversion efficiency and, therefore, serves as a useful optimization parameter to enhance performance [49,50]. Equations 9 and 10 provide two apparent routes: decreasing $J_{SRH}$ and $J_A$ and/or increasing $P_{esc}$ to increase $Q_e^{LED}(V_{OC})$. To evaluate which method best capitalizes on the benefits of photon recycling, we examine the relationship between $Q_e^{LED}(V_{OC})$ and $P_{esc}$ on $V_{OC}$ and $V_{MPP}$.

First, we consider how $Q_e^{LED}(V_{OC})$ and photovoltage are affected by decreasing $J_{SRH}$ (i.e. varying $k_1$ in Equation 7), for a fixed escape probability ($P_{esc}$ = 4.7%). Figure 4 shows $Q_e^{LED}(V_{OC})$ increases with decreasing $k_1$, resulting in voltage enhancements at open-circuit and MPP. We report a photon recycling threshold of $Q_e^{LED}(V_{OC})$ > ~0.3% and significant performance improvements for $Q_e^{LED}(V_{OC})$ = 10%, yielding $\Delta V_{OC}^{PR}$ = 36 mV and $\Delta V_{MPP}^{PR}$ = 9 mV (Figure 4). Recently, Liu *et al.* reported a $Q_e^{LED}(V_{OC})$ of 7.5±2.5% for CH$_3$NH$_3$PbI$_3$ devices achieving a $V_{OC}$ of 1.26 V. This experimental $V_{OC}$ is higher than the maximum achievable theoretical $V_{OC}$ in the radiative limit without photon recycling, indicating real-world performance enhancements due to photon recycling [45].

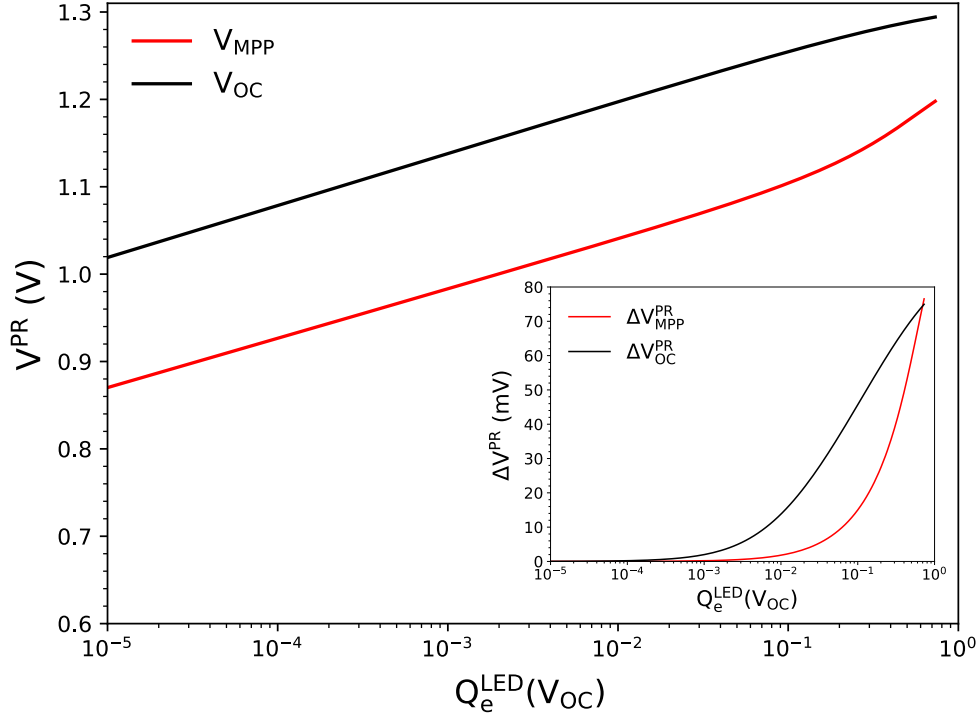

*Figure 4.* The voltage with photon recycling ($V^{PR}$) for $P_{esc}$ = 4.7% at MPP and open-circuit is shown as a function of $Q_e^{LED}(V_{OC})$, which, as non-radiative recombination decreases, approaches unity. Inset: $\Delta V^{PR}$ for $P_{esc}$ = 4.7% at MPP and open-circuit as a function of $Q_e^{LED}(V_{OC})$.

Second, we consider how $Q_e^{LED}(V_{OC})$ and photovoltage (see below) are affected through increasing $P_{esc}$ and decreasing $k_1$. We note that the photovoltage is proportional to the steady-state carrier density and is, therefore, an intuitive metric to compare across the multiple varying parameters. We calculate the carrier density using the law of mass action (Eq. 6) as a function of $k_1$ and $P_{esc}$.

Figure 5a shows that, at $P_{esc}$ = 4.7% and low $k_1$ values, photon recycling increases the steady-state carrier density by a factor of four, from 8.5×10$^{14}$ to 3.7×10$^{15}$ cm$^{-3}$ at MPP. This high carrier density results from additional generation associated with the reabsorption of trapped photons — up to 18-suns equivalent at open circuit and >1.4 equivalent suns at MPP (triple cation S6-S7, CH$_3$NH$_3$PbI$_3$ S19). Figure 5b shows that photon recycling allows $Q_e^{LED}(V_{OC})$ to exceed the escape probability, but it cannot reach 100% due to Auger recombination. Emission efficiencies larger than the escape probability result from multiple re-absorption events which re-randomize the photon propagation angle (S8 and S9).

Next, we consider scenarios in which $P_{esc}$ is changed without significantly impacting the material absorptivity function. Figure 5b and c show the carrier density and $Q_e^{LED}(V_{OC})$ for $P_{esc}$ = 9.4 and 14.1%. Here, the steady-state carrier density steadily decreases, while $Q_e^{LED}(V_{OC})$ approaches 90% with photon recycling due to a smaller contribution from Auger recombination at lower carrier densities. These results appear to counteract one another, as both a high steady-state carrier density and $Q_e^{LED}(V_{OC})$ are desired.

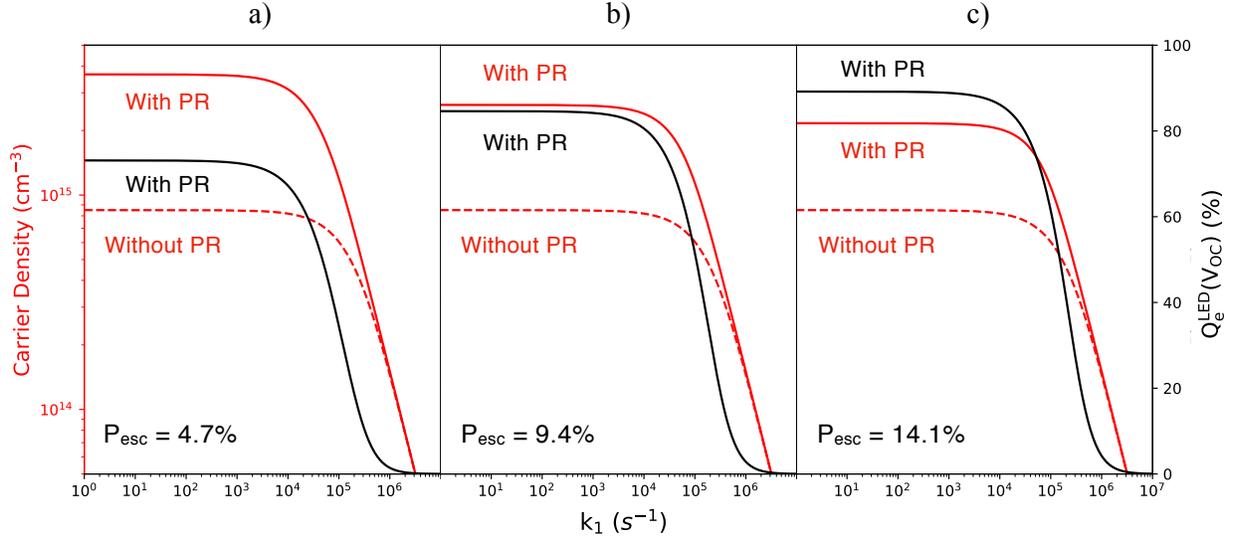

**Figure 5.** *The effect of photon recycling (PR) on the maximum power point (MPP) steady-state carrier density and $Q_e^{LED}(V_{OC})$ as a function of $k_1$ for a) $P_{esc}$ = 4.7%, b) 9.4%, and c) 14.1%.*

This observation raises the question as to whether solely increasing the escape probability can enhance device performance and, in particular, open-circuit voltage [15,16]. The traditional definition of the maximum achievable open-circuit voltage ($V_{OC}^{max}$) is expressed as a function of the external LED electroluminescence efficiency, as described in Equation 12 [36,47,51].

$$[12] \quad V_{OC}^{max} = V_{OC}^{rad} + \frac{kT}{q}\ln[Q_e^{LED}(V_{OC})]$$

Here, it appears that increasing $Q_e^{LED}(V_{OC})$ through enhancing the escape probability should allow $V_{OC}^{max}$ to approach $V_{OC}^{rad}$ — however, the implicit dependence of the radiative component ($V_{OC}^{rad}$) on $P_{esc}$ is often overlooked. This dependence becomes clear if we equate $J_0^{rad,ext}$ with the product of $J_0^{rad,int}$ and $P_{esc}$ (Eq. 13), where it can be seen that this term decreases as the escape probability increases [52].

$$[13] \quad V_{OC}^{rad} = \frac{kT}{q}\ln\left[\frac{J_{SC}}{J_0^{rad,ext}}\right] = \frac{kT}{q}\ln\left[\frac{J_{SC}}{P_{esc}J_0^{rad,int}}\right]$$

To better understand the competition between $V_{OC}^{rad}$ and $Q_e^{LED}(V_{OC})$ on $V_{OC}^{max}$, Figures 6a and b show the magnitude of these terms as a function of $P_{esc}$ and $k_1$. As $P_{esc}$ increases for a given $k_1$, $V_{OC}^{rad}$ decreases due to the enhanced light outcoupling, which increases the external radiative saturation current. Opposing this negative impact on $V_{OC}^{max}$ from $V_{OC}^{rad}$, $V_{OC}^{nonrad}$ also decreases with increasing $P_{esc}$, resulting in a smaller subtractive component from $V_{OC}^{max}$, as shown in Figure 6b. As $P_{esc}$ changes, the radiative and non-radiative terms vary in opposing directions.

Ultimately, $V_{OC}^{max}$ is dominated by the radiative dependence on $P_{esc}$, so $V_{OC}^{max}$ decreases monotonically with increasing $P_{esc}$ at a constant $k_1$ value (Fig. 6c) [16]. Thus, it is evident that simply increasing the outcoupling efficiency reduces output voltages due to the reduction in steady-state carrier density (c.f. Figure 5). We highlight that we only analyze the voltage in this simulation, and note that overall device performance may not track the changes in voltage if, for example, $J_{SC}$ changes as well.

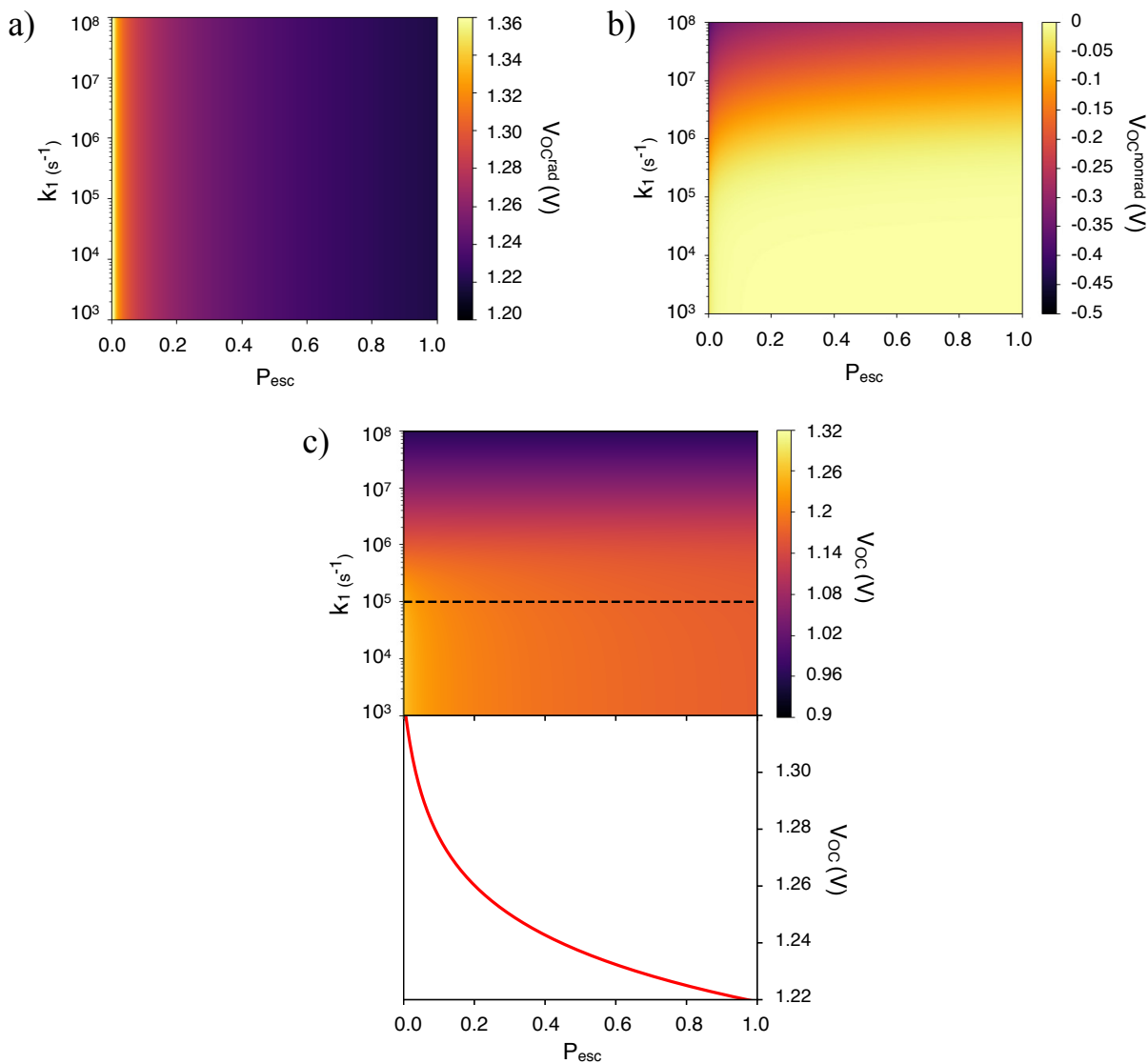

**Figure 6.** a) The $V_{OC}$ with photon recycling (PR) in the radiative limit ($V_{OC}^{rad}$) is shown along with b) the non-radiative subtractive effect on $V_{OC}^{max}$ ($V_{OC}^{nonrad}$). Combined, $V_{OC}^{rad} + V_{OC}^{nonrad}$ yield c) the total $V_{OC}^{max}$ as a function of $k_1$ and $P_{esc}$, with a dashed line at $k_1 = 1\times10^5$ s$^{-1}$ showing that increasing $P_{esc}$ for a fixed $Q_i^{lum}$ decreases $V_{OC}$.

In summary, we have presented a rigorous method for evaluating the extent of and benefits from photon recycling in emerging perovskite absorbers by exploring device performance limits using experimentally determined optical constants and absorption for triple cation films. This analysis investigates the effect of photon recycling on both $V_{OC}$ and operationally-relevant maximum power point (MPP) parameters, both in the radiative limit and with non-radiative recombination. Our simulations provide a framework for evaluating the improvements attributable to photon recycling in standard current-voltage measurements. This analysis reveals that perovskite devices demonstrating voltage deficits of <0.38 V [29,45] already benefit from photon recycling. This would mean that high-quality devices fabricated today may be further improved by reducing non-radiative recombination and/or modifying the escape probability to harness the benefits of photon recycling.

With recycling, photons waveguided within the film can be re-absorbed and re-emitted in the escape cone, allowing $Q_e^{LED}(V_{OC})$ to approach the intrinsic limit while maintaining a high steady-state carrier density. If $Q_e^{LED}(V_{OC})$ is enhanced only by increasing $P_{esc}$, the steady-state carrier density subsequently decreases, resulting in a lower $V_{OC}$. For triple cation films, enhancements in $V_{OC}$ and $V_{MPP}$ are observed for $k_1 < 2 \times 10^6$ s$^{-1}$ ($\tau_1 > 500$ ns) and $k_1 < 7 \times 10^5$ s$^{-1}$ ($\tau_1 > 1430$ ns), respectively, while, for $k_1 < 1 \times 10^4$ s$^{-1}$ ($\tau_1 > 100$ μs), further performance improvements become negligible. Our analysis identifies a target non-radiative recombination rate for perovskite films of $k_1 = 1 \times 10^4$ s$^{-1}$. Below this threshold, the steady-state carrier density plateaus at 4× the density without photon recycling. In theory, a perovskite film reaching this lower bound of $k_1 = 1 \times 10^4$ s$^{-1}$ ($P_{esc}$ = 4.7%) can achieve a 74 mV increase in $V_{OC}$, 73 mV improvement in $V_{MPP}$, 0.3% absolute increase in fill factor, and 1.79% increase in PCE — due solely to photon recycling.

We note that the model used to simulate J-V curves in this study represents an ideal case and sets an upper limit for the target non-radiative recombination rate constants. For example, perovskite material quality will likely need to be even better than these targets, as charge transport layers in devices introduce new pathways for interfacial recombination. Passivation methods and surface modifiers that reduce the number of defects at the interfaces and lead to favorable band alignment will be critical in minimizing non-radiative loss to fully harness photon recycling [18,24,53]. Toward the development of new perovskite formulations and device architectures, this analysis provides clear material quality targets and device performance limits for evaluating photon recycling in next-generation perovskite solar cells.

**Acknowledgments**

This work is supported by the TATA-MIT GridEdge Solar Research program. This material is based upon work supported by the National Science Foundation Graduate Research Fellowship under Grant No. (1122374). We thank Dak Benjia Dou for his guidance in the preparation of the triple cation perovskite thin films; Nina Hong, from J.A. Woollam Co., Inc., for her technical support analyzing the ellipsometry data; and Luis Pazos-Outón and Thomas Mahony for valuable discussions.

# Supporting Information

## State-of-the-Art Perovskite Solar Cells Benefit from Photon Recycling at Maximum Power Point

Roberto Brenes[1,3, †], Madeleine Laitz[1,3, †], Joel Jean[1,2], Dane W. deQuilettes[3], Vladimir Bulović[1,3*]

[1]Department of Electrical Engineering and Computer Science, Massachusetts Institute of Technology, 77 Massachusetts Avenue, Cambridge, Massachusetts 02139, USA
[2]Swift Solar Inc., Golden, Colorado 80401, USA
[3]Research Laboratory of Electronics, Massachusetts Institute of Technology, 77 Massachusetts Avenue, Cambridge, Massachusetts 02139, USA

†Equal Contribution

*Corresponding Author: bulovic@mit.edu

*rbrenes@mit.edu, mlaitz@mit.edu*

- Experimental Methods
- Data Analysis
- Model Assumptions
- Triple Cation Supplementary Figures
- CH$_3$NH$_3$PbI$_3$ Supplementary Figures
- References

# Experimental Methods

**Materials**

Formamidinium iodide (FAI) and methylammonium bromide (MABr) were purchased from GreatCell Solar. Lead iodide ($PbI_2$) and lead bromide ($PbBr_2$) were obtained from TCI Chemicals. Cesium iodide (CsI), *N*,*N*-dimethylformamide (DMF), dimethyl sulfoxide (DMSO) and all other chemicals were sourced from Sigma Aldrich unless otherwise stated.

**Sample Preparation**

Glass, fused silica, and p-type silicon substrates were washed sequentially with soap (2% Hellmanex in water), de-ionized water, acetone, and isopropanol, and finally treated under UV-Ozone for 30 min.

Thin films of $Cs_{0.05}(MA_{0.17}FA_{0.83})_{0.95}Pb(I_{0.83}Br_{0.17})_3$ (triple cation) were solution processed following the procedure outlined by Saliba *et al* [1]. The precursor solution was prepared in a nitrogen-filled glovebox by dissolving 1 M FAI, 0.2 M MABr, 1.1 M $PbI_2$, 0.2 M $PbBr_2$, and 0.05 M CsI in a 4:1 volume ratio of DMF to DMSO. The solution was spin coated in a dry-air box with a two-step program, with the first step at 1,000 rpm for 10 s with a 1000 rpm/s ramp followed by a step at 6000 rpm for 20 s with a 6000 rpm/s ramp. 110 µl of chlorobenzene was added 5 s before the spin procedure ended. The films were then annealed at 100°C for 1 hour. Samples were then stored in the dark in a nitrogen-filled glovebox until use.

Thin films deposited on top of fused silica were used for PDS and UV-Vis measurements, while films deposited on top of silicon were used for ellipsometry measurements.

**Photothermal Deflection Spectroscopy (PDS)**

PDS measurements were performed using a custom system optimized for the near-infrared (**S1**) [2]. The pump beam consists of a 300 W Xe arc lamp chopped at 10 Hz, a dual-grating monochromator (300 lines per mm) with 1 mm slits (15 nm output FWHM), a periscope, and achromatic lenses to collimate and focus the beam. The illuminated spot size at the sample is approximately 3.5 mm wide × 0.4 mm tall, corresponding to a monochromatic intensity of 50–750 mW cm$^{-2}$.

The sample is secured using a custom holder in a standard 10 mm quartz cuvette. Perfluorohexane (Acros Organics Fluorinert FC-72) filtered with 0.02 μm PTFE is used as a deflection medium.

The probe beam consists of a 658 nm, 40 mW temperature-controlled laser diode, an anamorphic prism pair, spatial filter, and iris to circularize the beam and isolate the fundamental mode; a bandpass filter to eliminate scattered light; and a quadrant detector with built-in transimpedance amplifier. A DAQ is used to drive the chopper and acquire the quad detector signal. AC lock-in detection is performed using custom LabView software with a low-pass filter cutoff frequency of 0.5 Hz. The entire PDS system is assembled on an optical table to minimize vibrations and enclosed in a box to mitigate stray light and air flow.

**Variable Angle Spectroscopic Ellipsometry**

Spectroscopic ellipsometry was performed using a variable angle spectroscopic ellipsometer (Woollam) at 65°, 70°, and 75° angles of incidence. Ellipsometry data were fitted to obtain film optical constants and thicknesses used in PDS data analysis.

**UV-Vis Reflectivity Measurements**

Reflection spectrophotometry was performed with light incident from the film side using an Agilent Cary 5000 dual-beam UV–vis–NIR spectrophotometer. Specular reflectance was collected at an incident angle of 8°. A 3 mm round aperture was used for all measurements.

# Data Analysis

**PDS Data Analysis**

Acquired PDS data for the perovskite samples was normalized with respect to the pump power spectrum. The data was then re-normalized to one at a highly absorbing wavelength (400 nm) and divided by the PDS spectrum of a black reference sample to obtain a relative absorptance spectrum. Absolute absorptance was then obtained by scaling the data by (1-*R*), where *R* is the reflectivity of the perovskite sample at a highly absorbing wavelength obtained from spectrophotometry measurements. To determine the absorption coefficient from the absolute absorptance spectrum, we followed the procedure developed by Ritter and Weisser utilizing thicknesses and optical constants for fused silica and the perovskite absorber layer determined from ellipsometry [3].

**Corrected Absorption Coefficient Spectrum**

The absorption coefficient spectrum determined from ellipsometry was used as the basis data set. Since PDS measurements provide increased sensitivity by multiple orders of magnitude, the absorption coefficient spectrum obtained from PDS measurements was used for the band edge. The PDS data was shifted by 20 meV to match with the absorption coefficient spectrum from ellipsometry. The Urbach expression $\alpha = \alpha_0 \exp(\frac{E-E_0}{E_U})$, where $\alpha$ is the absorption coefficient, $\alpha_0$ is an absorption coefficient scaling factor, $E$ is the photon energy, $E_0$ is an energy offset and $E_U$ is the Urbach energy was fitted to the band edge to extrapolate the absorption coefficient spectrum beyond the noise floor of the PDS measurements. We obtained a fitted value of $E_U = 15$ meV, consistent with previous reports for other perovskite absorber materials [4,5].

# Model Assumptions

**Critical Mobility**

In the radiative Shockley-Queisser performance limit, it is assumed that all photogenerated carriers are entirely collected, which would require charge carriers to possess extremely high mobilities. Mattheis *et al*. examined the effect of mobility on idealized photovoltaic devices to determine whether, in relevant materials, there is a PCE ceiling before the SQ limit based on fundamental mobility limitations, even in the radiative limit [6]. A critical mobility was defined based on a reference mobility ($\mu_{ref}$) for each material, with $\mu_{ref}$ defined in Eq. 1:

$$[1] \quad \mu_{ref} = \frac{q N_A \Phi_{bb}^{E_g}}{kT \alpha_0 n_i^2}$$

where $q$ is the electron charge, $N_A$ the doping density, $\Phi_{bb}^{E_g} = \int_{E_g}^{\infty} \phi_{bb}(E) dE$ is the integrated black body spectrum $\phi_{bb}(E)$ is the black body spectral photon flux, $k$ is the Boltzmann constant, $T$ is the temperature of the material, and $\alpha_0$ is the average absorption coefficient across the AM1.5 spectrum. Mattheis *et al.* showed that the SQ radiative limit can be achieved if the material's mobility was two orders of magnitude larger than $\mu_{ref}$ and thus the limiting efficiency was not limited by mobility. Using typical reported doping concentration of $N_A$ = 1x10$^{16}$ cm$^{-3}$ for MAPbI$_3$ films, an intrinsic carrier concentration of 2.74x10$^5$ cm$^{-3}$, and the average absorption coefficient across the AM1.5 spectrum for MAPbI$_3$, $\mu_{ref}$ = 0.013 cm$^2$V$^{-1}$s$^{-1}$ [7,8]. Using typically reported mobilities for MAPbI$_3$ films, $\mu_{MAPbI3}$ = 3 cm$^2$V$^{-1}$s$^{-1}$ = 225*$\mu_{ref}$ [9]. The typical mobility of MAPbI$_3$ is more than two orders of magnitude greater than the reference mobility, and thus we will assume that this analysis is not limited by mobility, and that perfect charge collection can be assumed even when investigating the effect of non-zero non-radiative recombination on JV characteristics. The model can then be extended to determine the impact of photon recycling on device performance when Shockley-Read Hall ($k_{nr}$) and Auger ($k_A$) recombination rates are non-zero, both at open circuit and at the maximum power point of the device.

**Band Gap of Triple Cation Thin Films**

The band gap for Cs$_{0.05}$(MA$_{0.17}$FA$_{0.83}$)$_{0.95}$Pb(I$_{0.83}$Br$_{0.17}$)$_3$ used in the simulations is as reported by Soufiani *et al.* for planar Cs$_{0.05}$(MA$_{0.17}$FA$_{0.83}$)$_{0.95}$Pb(I$_{0.83}$Br$_{0.17}$)$_3$ films at 2K: E$_g$ = 1.593 eV [10]. The second-order radiative recombination constant used for the simulations in this work was reported by Kumar et al., with a corresponding steady-state carrier density of 2.3x10$^5$ cm$^{-3}$.

**Calculating Total, Short-Circuit, and Radiative Saturation Currents**

$J_{total}$, $J_{SC}$, and the radiative saturation current (internal and external) are calculated assuming angle independence of irradiation, optical constants, and absorption. Furthermore, when calculating $J_{sc}$ for these simulations, we assume perfect incoupling due to the low short-circuit current deficit between theoretical and experimental reports ($J_{sc}^{theoretical}$ = 25.8 mA cm$^{-2}$ and $J_{sc}^{experimental}$ = 24.5 mA cm$^{-2}$ [11]).

**V$_{OC}$ Enhancement Due to Photon Recycling**

The expression for the enhancement in V$_{OC}$ solely due to photon recycling is as derived by Abebe *et al.* (Eq. 2) [12].

$$[2] \quad \Delta V_{OC}^{PR} = \frac{kT_c}{q} \ln\left[\frac{1}{1-\text{PLQE}_{int}(1-P_{esc})}\right]$$

**Internal Suns**

The internal photon population corresponds to the addition of incident photons and the photons generated by radiative recombination that are waveguided within the film. This can be expressed in terms of the short circuit current and saturation currents which, when normalized with respect to the short circuit current, provide an equivalent number of internal suns (Eq. 3).

$$[3] \quad Suns = \frac{J_0^{rad,int}(1-P_{esc})+J_{sc}-J_{SRH}-J_A}{J_{sc}}$$

# Supplementary Figures: Triple Cation

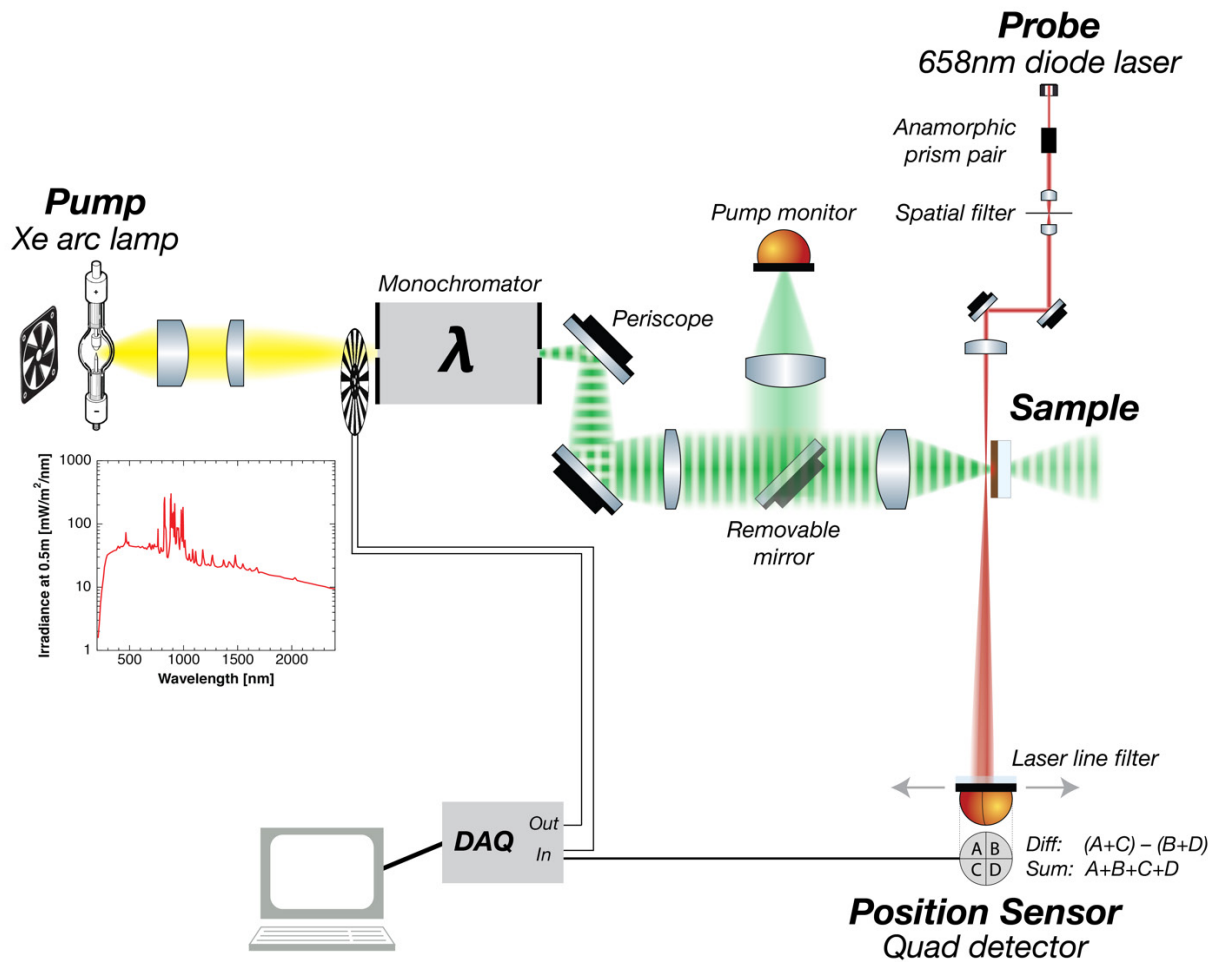

*Figure S1.* *Schematic of photothermal deflection spectroscopy with Xe arc lamp pump and 658 nm diode laser probe* [2].

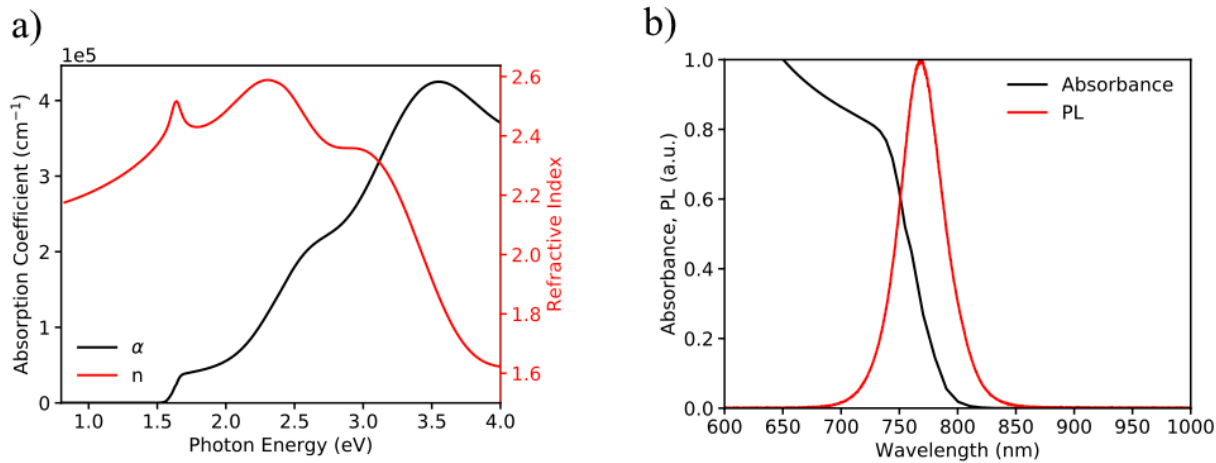

*Figure S2.* a) Absorption coefficient (black trace) and index of refraction (red trace) of triple cation films as a function of photon energy. b) Absorption spectrum (black trace) and photoluminescence (PL) spectrum (red trace) of triple cation films. The overlap between the two spectra is due to the low Stokes shift in the material.

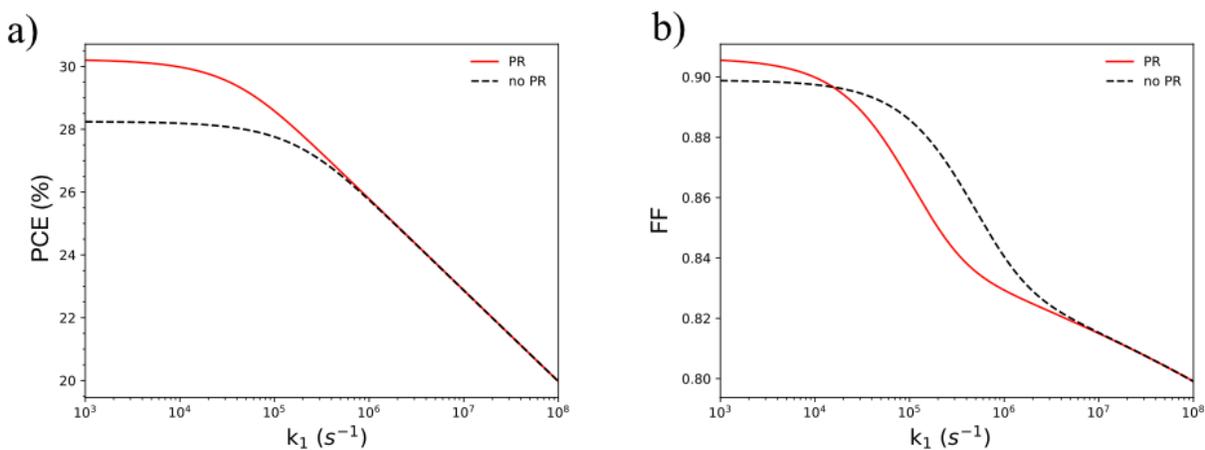

*Figure S3.* a) Power conversion efficiency (PCE) and b) fill-factor (FF) as a function of non-radiative recombination rate ($k_1$) with and without photon recycling. With photon recycling, the fill-factor initially decreases because of the larger fraction of non-radiative recombination at the maximum-power-point. Ultimately, when $k_1$ decreases below ~$2 \times 10^4$ $s^{-1}$, radiative recombination outcompetes non-radiative recombination and the fill-factor with photon recycling is greater than the fill-factor without photon recycling.

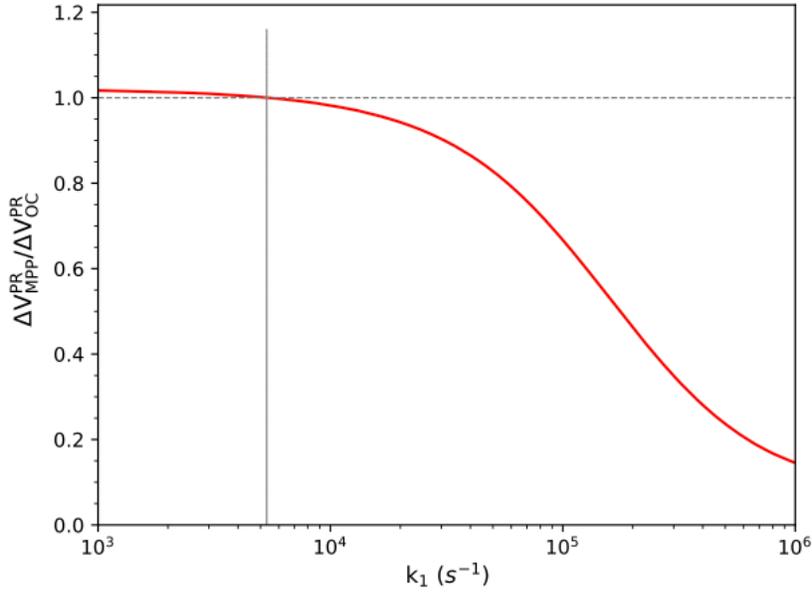

*Figure S4.* The increase in voltage at the maximum-power-point (MPP) due to photon recycling ($\Delta V_{MPP}^{PR}$) divided by the increase in voltage at open-circuit due to photon recycling ($\Delta V_{OC}^{PR}$) is shown as a function of $k_1$. As $k_1$ decreases, the fraction exceeds 1, indicating that, at $k_1$ ~$5\times10^3$ s$^{-1}$, $\Delta V_{MPP}^{PR}$ increases faster than $\Delta V_{OC}^{PR}$.

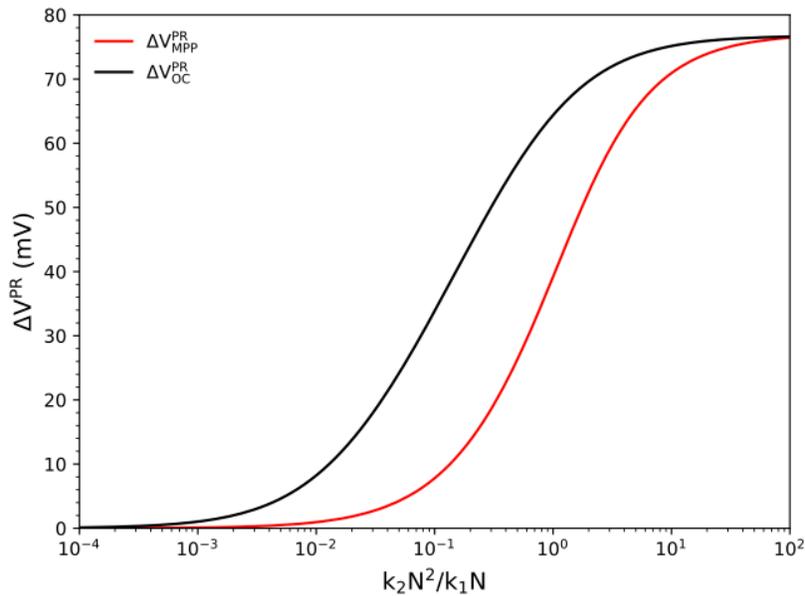

*Figure S5.* The ratio of the second order radiative recombination constant multiplied by the carrier density squared to the first order non-radiative recombination rate constant multiplied by the carrier density. When $k_2N^2/k_1N$ = ~10, there is a ~70 mV improvement in $V_{MPP}$.

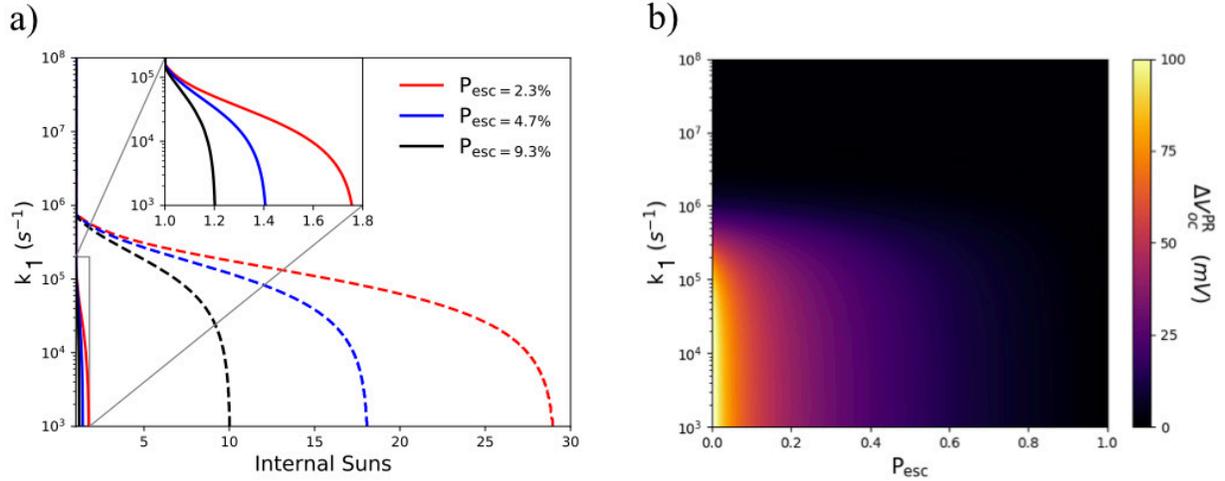

**Figure S6.** a) Internal suns as a function of $k_1$ at maximum power point (inset, solid lines) and at $V_{OC}$ (dashed lines) for varying probability of escape ($P_{esc}$). b) Increase in open-circuit voltage due to photon recycling ($\Delta V_{OC}^{PR}$) as a function of $k_1$ and $P_{esc}$.

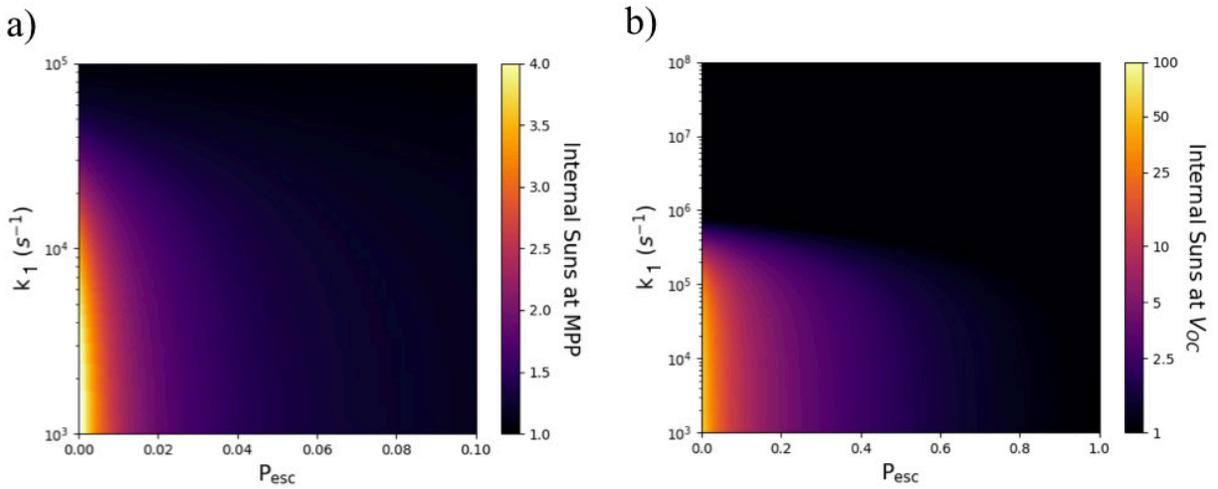

**Figure S7.** Internal suns at a) MPP and b) open-circuit as a function of $k_1$ and $P_{esc}$. Internal suns at MPP is shown on reduced axes from $P_{esc} = 0 - 10\%$ and $k_1 = 10^3 - 10^5$ s$^{-1}$, whereas the internal suns at open-circuit is shown from $P_{esc} = 0 - 100\%$ and $k_1 = 10^3 - 10^8$ s$^{-1}$.

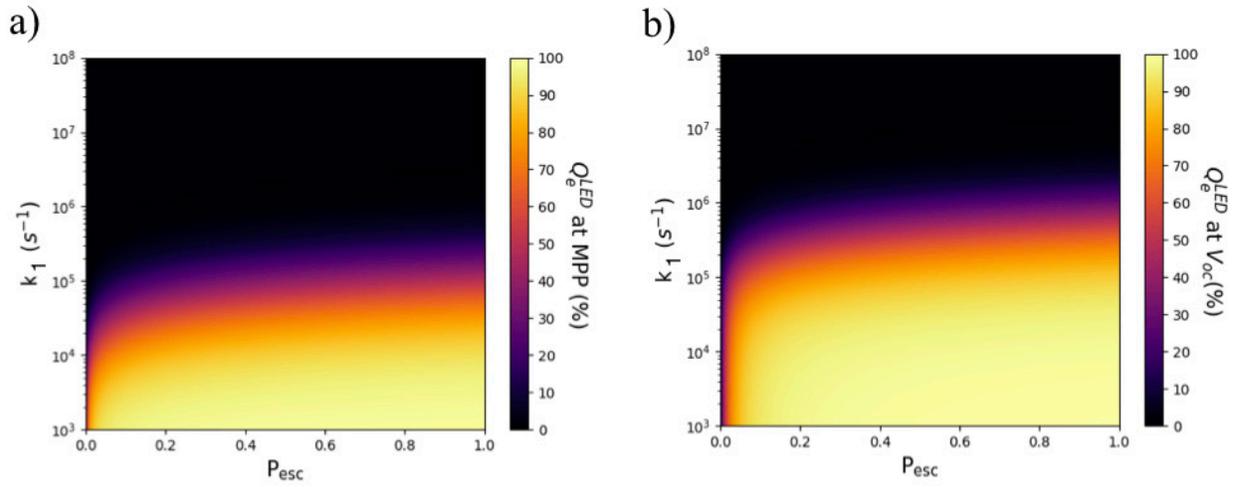

**Figure S8.** *The external electroluminescence efficiency ($Q_e^{\text{LED}}$) at a) MPP and b) open-circuit as a function of $k_1$ and $P_{esc}$.*

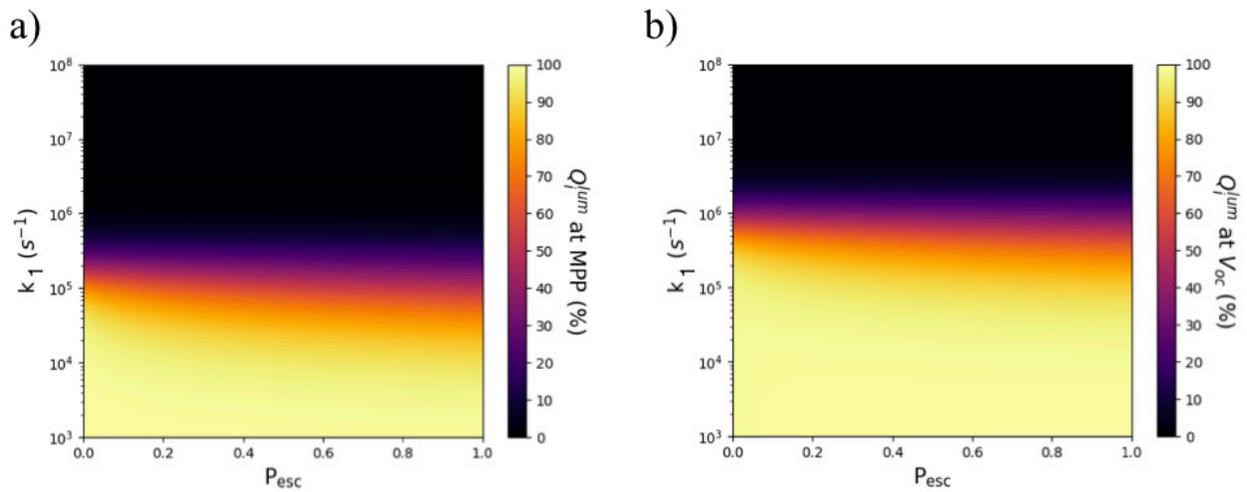

**Figure S9.** *The internal luminescence efficiency ($Q_i^{\text{lum}}$) at a) MPP and b) open-circuit as a function of $k_1$ and $P_{esc}$.*

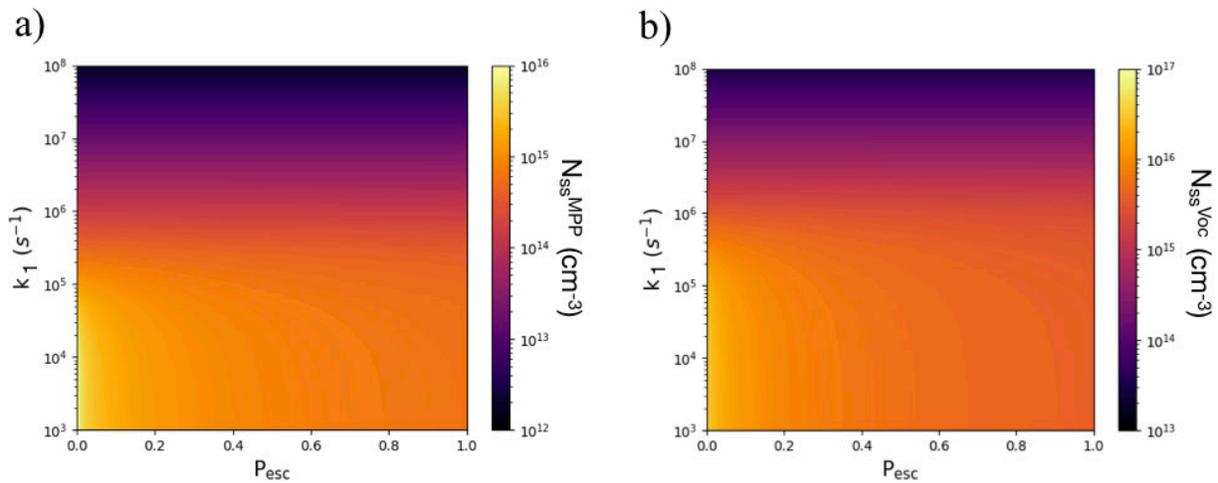

*Figure S10.* The steady-state carrier density ($N_{ss}$) at a) MPP and b) open-circuit as a function of $k_1$ and $P_{esc}$.

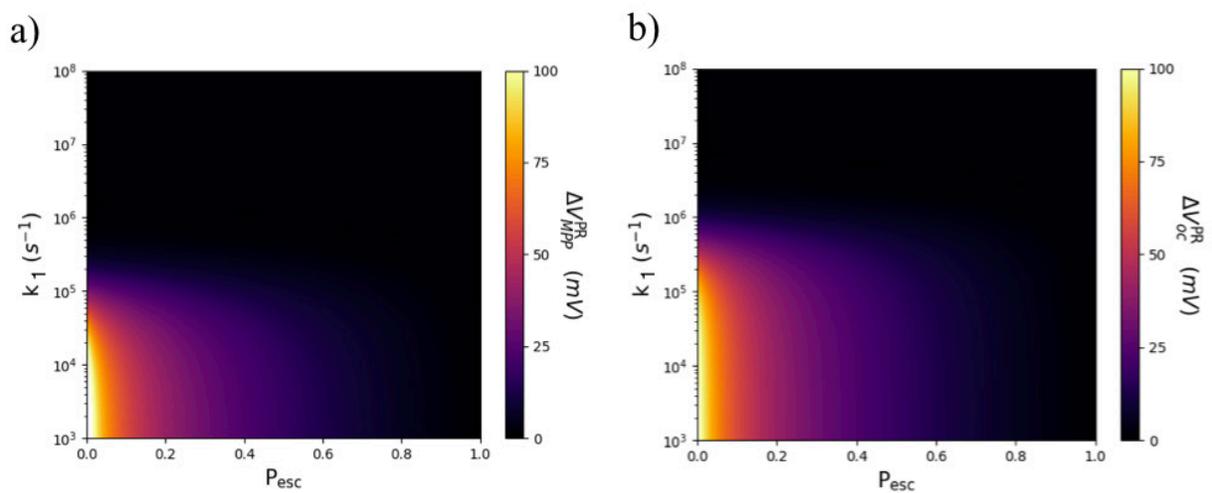

*Figure S11.* Increase in voltage due to photon recycling ($\Delta V^{PR}$) at a) MPP and b) open-circuit as a function of $k_1$ and $P_{esc}$.

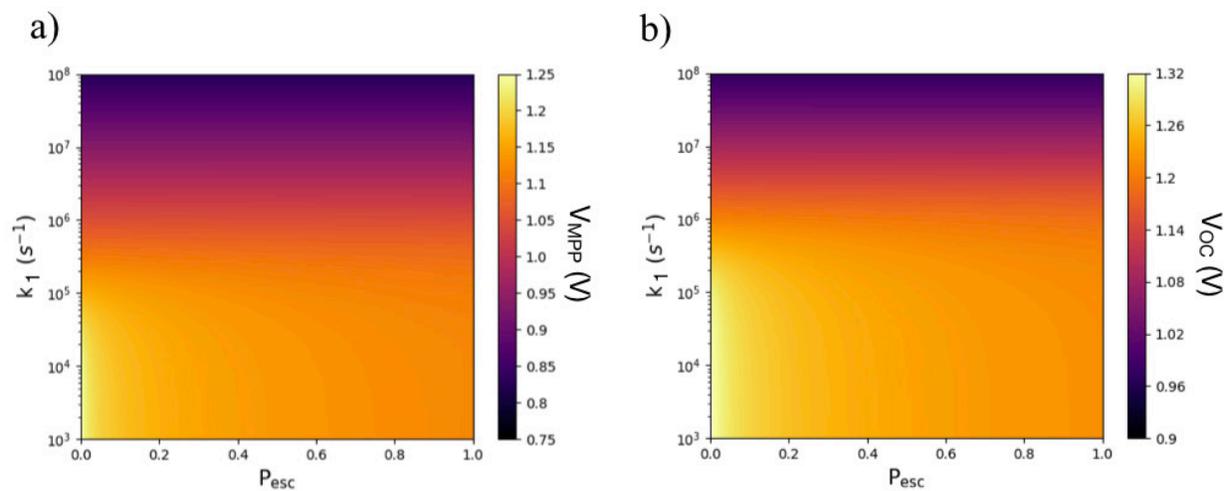

***Figure S12.*** *Voltage at a) MPP and b) open-circuit as a function of $k_1$ and $P_{esc}$.*

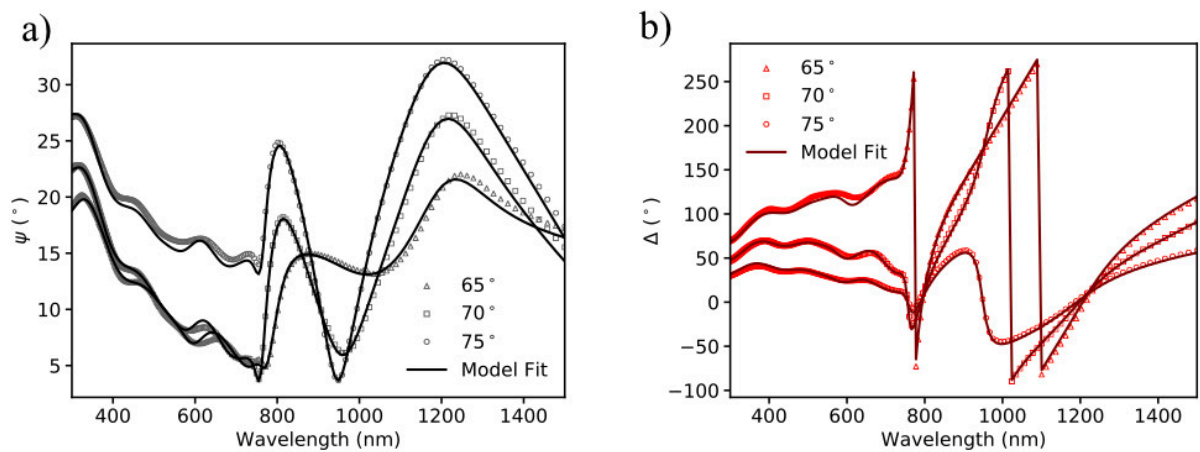

***Figure S13.*** *a) $\psi$ and b) $\Delta$ variable angle spectroscopic ellipsometry (VASE) data acquired at incident angles 65° (triangles), 70° (squares) and 75° (circles).*

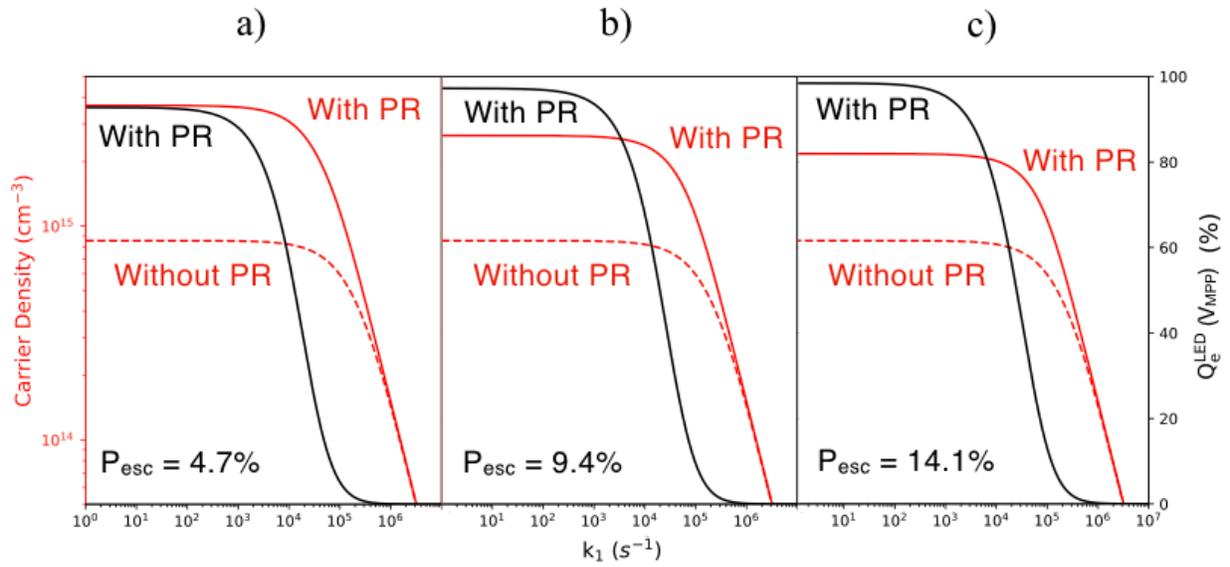

**Figure S13.** *The effect of PR on MPP steady-state carrier density and $Q_e^{LED}(V_{MPP})$ (calculated with an injection current achieved at a voltage bias of $V_{MPP}$) as a function of $k_1$ for a) $P_{esc}$ = 4.7%, b) 9.4%, and c) 14.1%.*

## Supplementary Figures: CH₃NH₃PbI₃

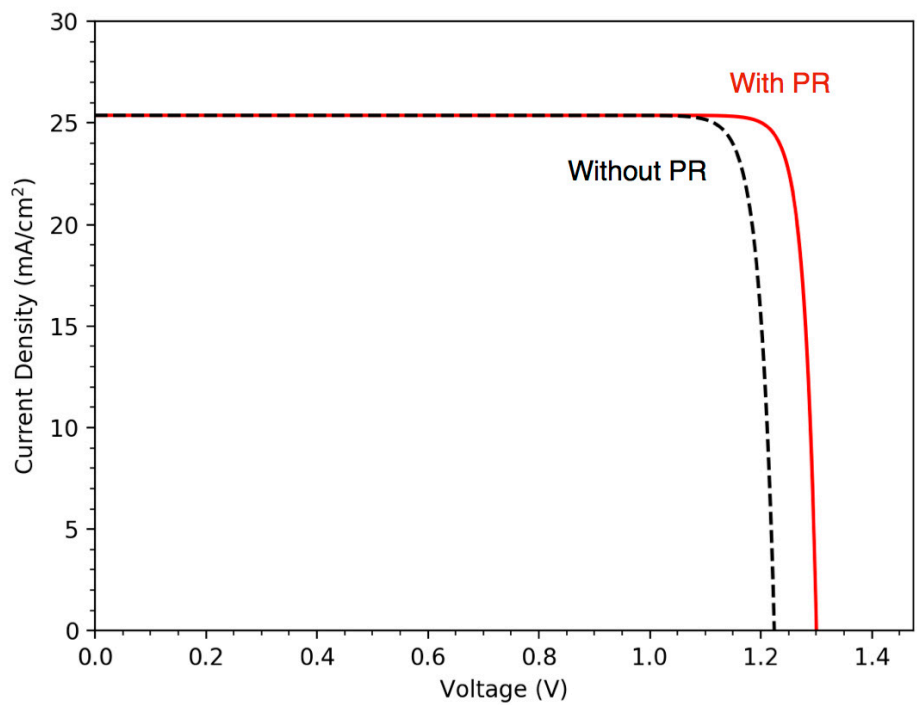

***Figure S14.*** *J-V characteristics in the radiative limit (no non-radiative recombination) with (red curve) and without (black dashed curve) photon recycling.*

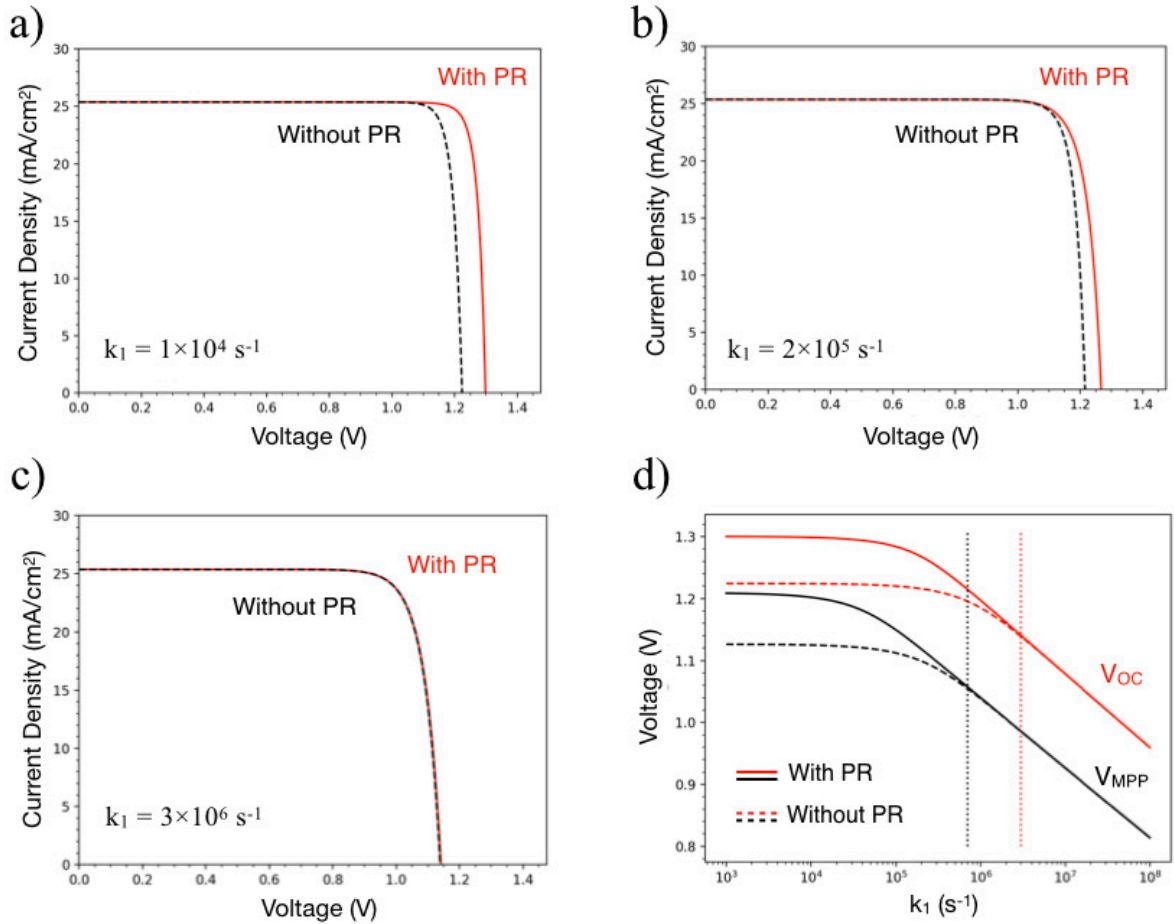

*Figure S15.* Simulated J-V curves (MAPbI$_3$) with and without photon recycling for $k_1$ values of (a) 1×10$^4$ s$^{-1}$, (b) 2×10$^5$ s$^{-1}$, and (c) 3×10$^6$ s$^{-1}$, all with fixed values for the radiative ($k_2$ = 1.14×10$^{-10}$ cm$^3$ s$^{-1}$) and Auger ($k_3$ = 1×10$^{-28}$ cm$^6$ s$^{-1}$) rate constants. (d) $V_{OC}$ (red lines) and $V_{MPP}$ (black lines) are shown as a function of $k_1$, revealing differences in the onset of performance improvements due to photon recycling. Dotted vertical red and black lines indicate $k_1$ thresholds (2.2×10$^6$ s$^{-1}$ and 7×10$^5$ s$^{-1}$, respectively) below which photon recycling improves performance at open-circuit and at MPP, respectively.

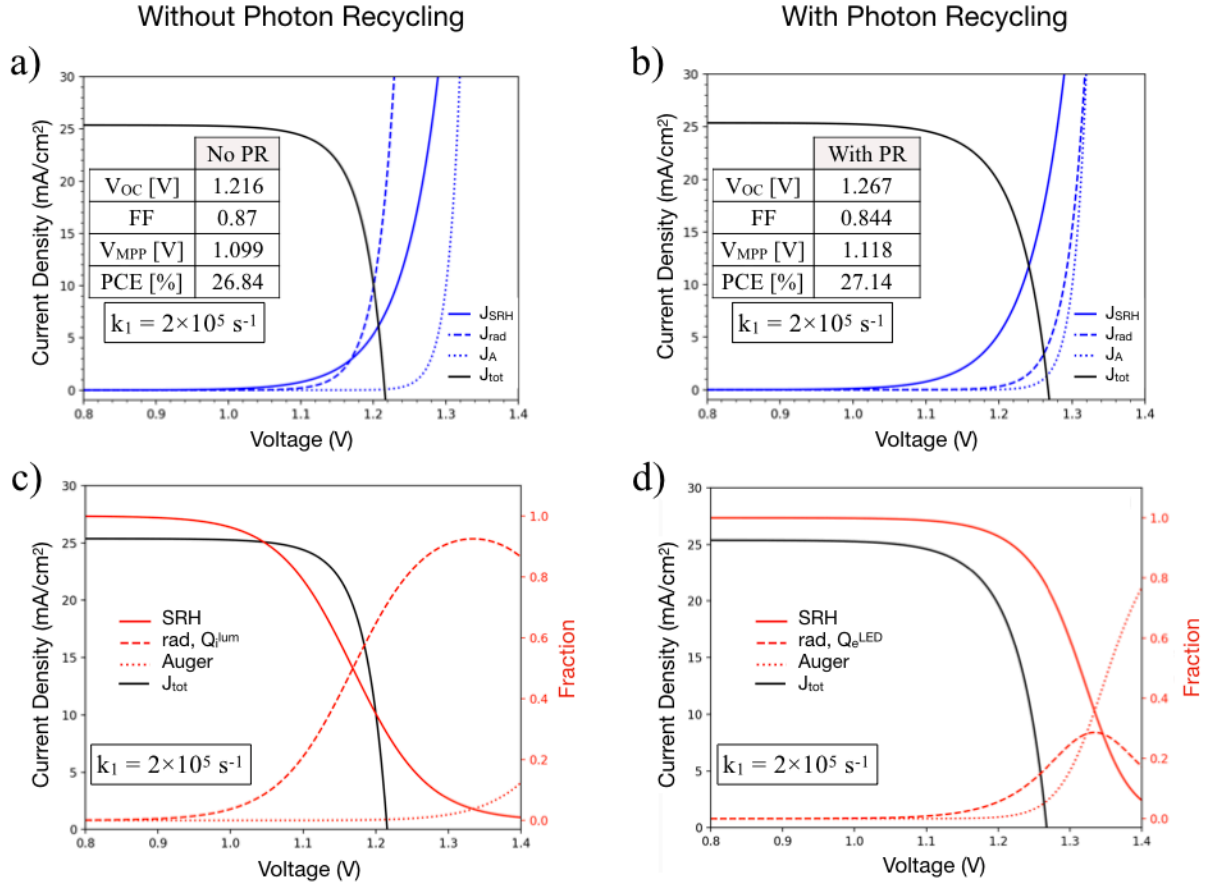

**Figure S16.** The J-V curve for $k_1 = 2 \times 10^5$ s$^{-1}$ (a, c) without and (b, d) with photon recycling (black traces) is depicted along with the magnitude of the three recombination currents (SRH, radiative, and Auger) as a function of voltage (blue traces). (c,d) The fraction of total recombination current due to SRH, radiative, and Auger recombination is shown at each voltage (red traces) (c) without and (d) with photon recycling. The fraction of radiative recombination as a function of voltage with and without photon recycling is equivalent to $Q_e^{\text{LED}}$ and $Q_i^{\text{lum}}$, respectively.

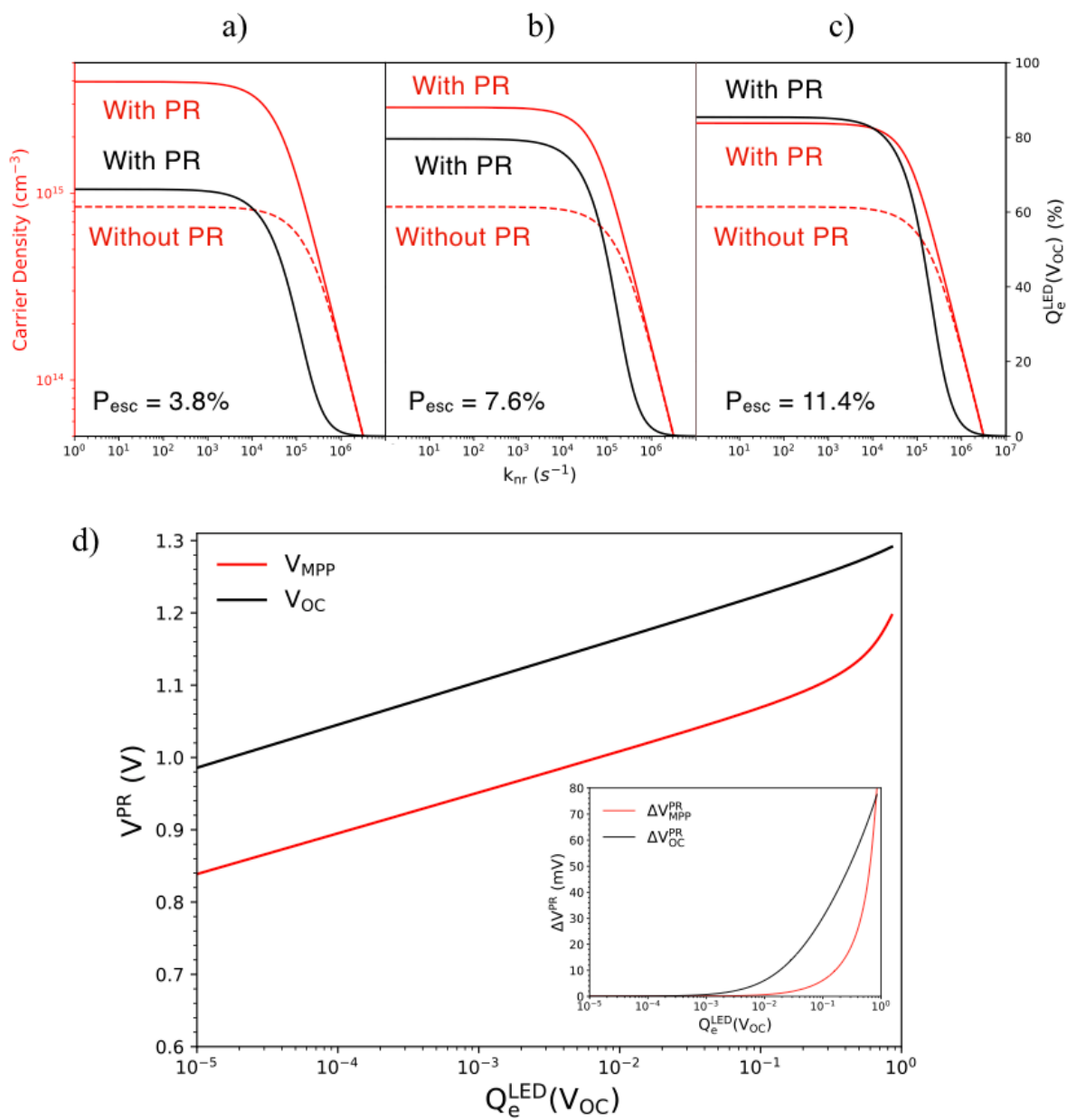

*Figure S17.* The effect of photon recycling on steady-state carrier density and $Q_e^{\text{LED}}(V_{\text{OC}})$ as a function of non-radiative recombination rate for a) $P_{esc}$ = 3.8% (probability of escape for MAPbI$_3$), b) $P_{esc}$ = 7.8%, and c) $P_{esc}$ = 11.4%. d) $\Delta V^{PR}$ for $P_{esc}$ = 3.8% at maximum-power-point and open-circuit is shown as a function of $Q_e^{\text{LED}}(V_{\text{OC}})$, which, as $k_1$ decreases, approaches 90%.

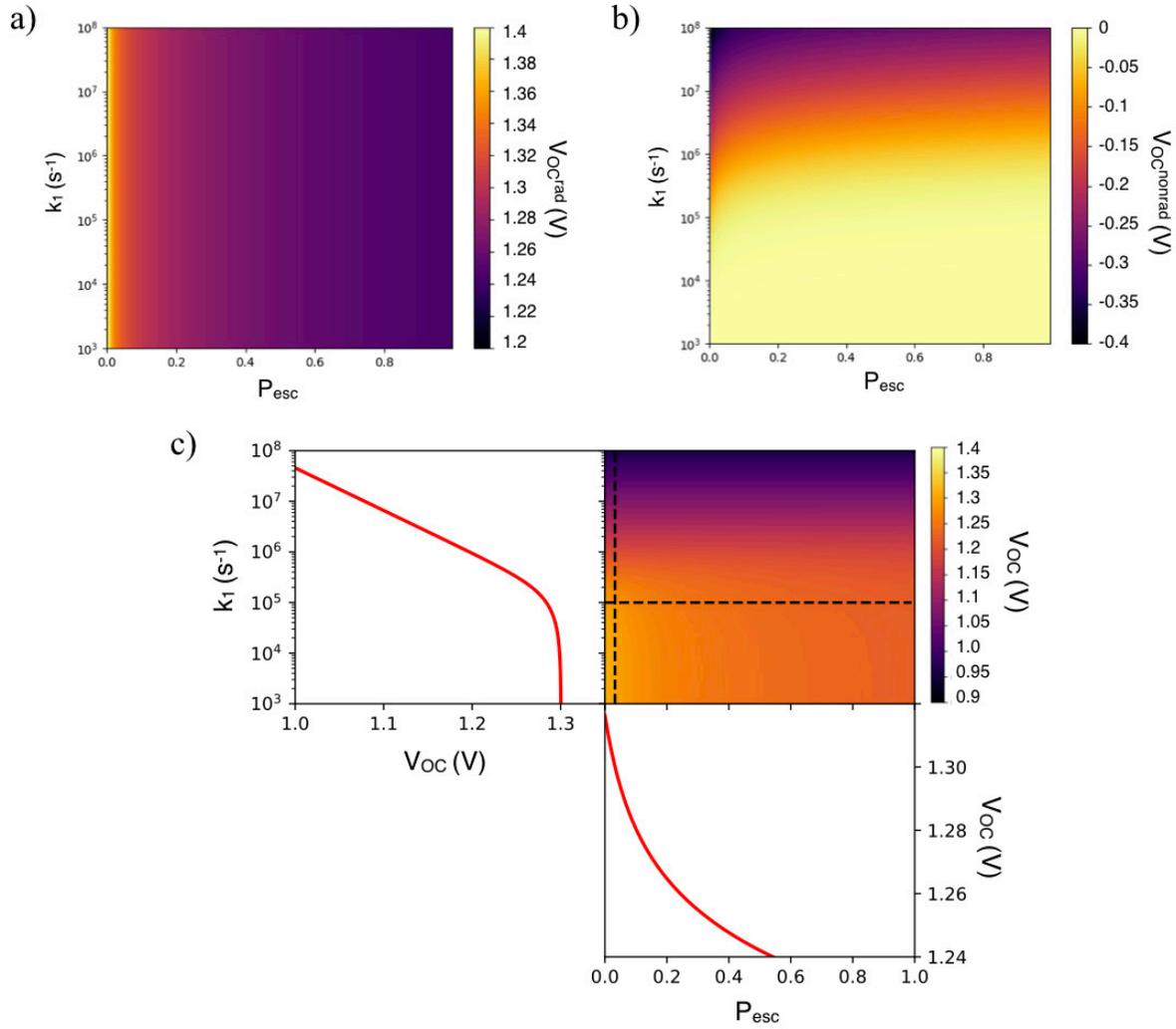

***Figure S18.*** *The $V_{OC}$ with photon recycling in the radiative limit ($V_{OC}^{rad}$) is shown along with b) the non-radiative subtractive effect on $V_{OC}^{max}$ ($V_{OC}^{nonrad}$). Combined, $V_{OC}^{rad} + V_{OC}^{nonrad}$ yield c) the total $V_{OC}^{max}$ as a function of $k_1$ and $P_{esc}$ with dashed vertical and horizontal lines indicating $P_{esc}$ = 3.8% (for MAPbI₃) and $k_1$ = 1×10⁵ s⁻¹, respectively. Increasing $P_{esc}$ for a fixed $Q_i^{lum}$ decreases $V_{OC}$.*

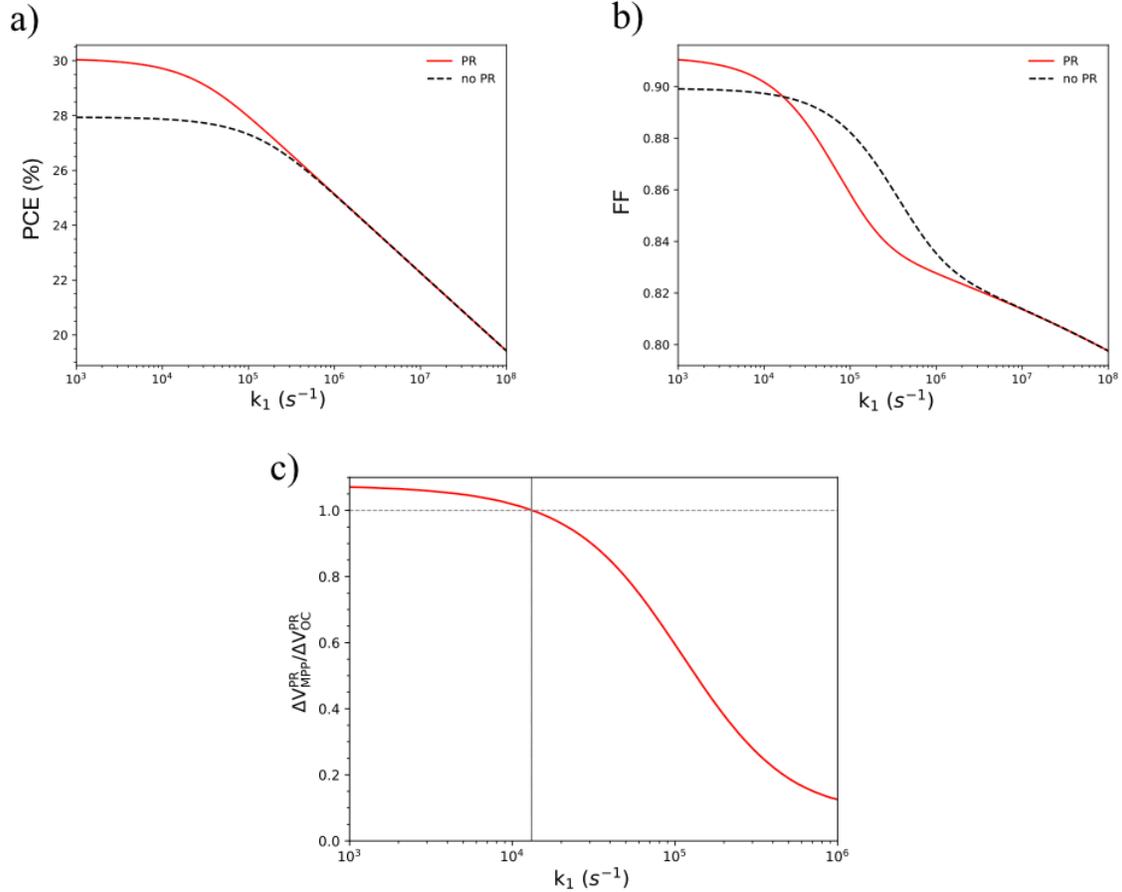

**Figure S19.** a) PCE and b) FF as a function $k_1$ with (red trace) and without (dotted black trace) photon recycling. When $k_1 < \sim 1.5 \times 10^4$ s$^{-1}$, radiative recombination outcompetes non-radiative recombination and the FF with PR is greater than the FF without PR. c) $\Delta V_{MPP}^{PR}/\Delta V_{OC}^{PR}$ is shown as a function of $k_1$. At $k_1 \sim 1.5 \times 10^4$ s$^{-1}$, $\Delta V_{MPP}^{PR}$ increases faster than $\Delta V_{OC}^{PR}$.

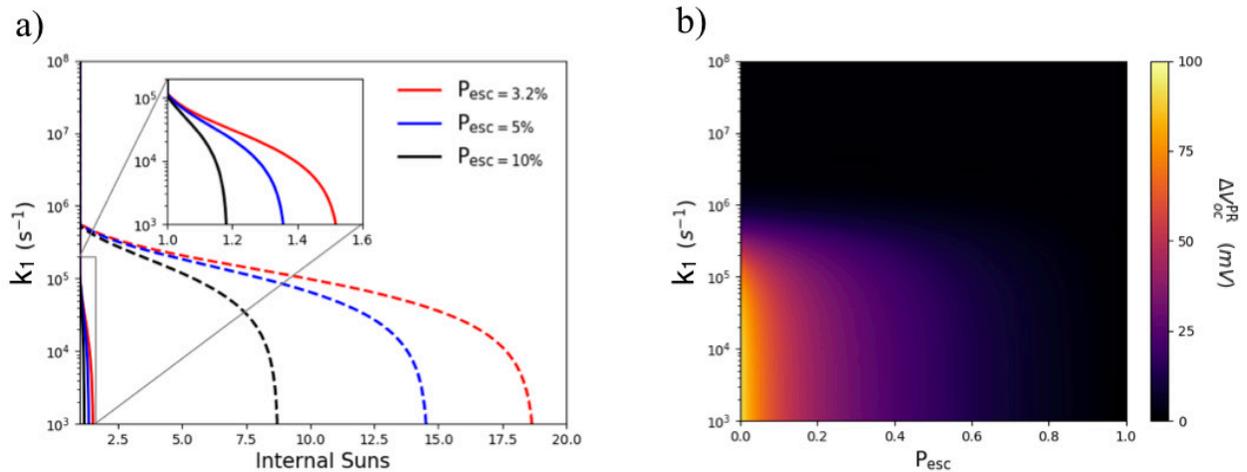

**Figure S20.** a) Internal suns as a function of $k_1$ at MPP (inset, solid lines) and at $V_{OC}$ (dashed lines) for varying $P_{esc}$. b) $V_{OC}^{PR}$ as a function of $k_1$ and $P_{esc}$.

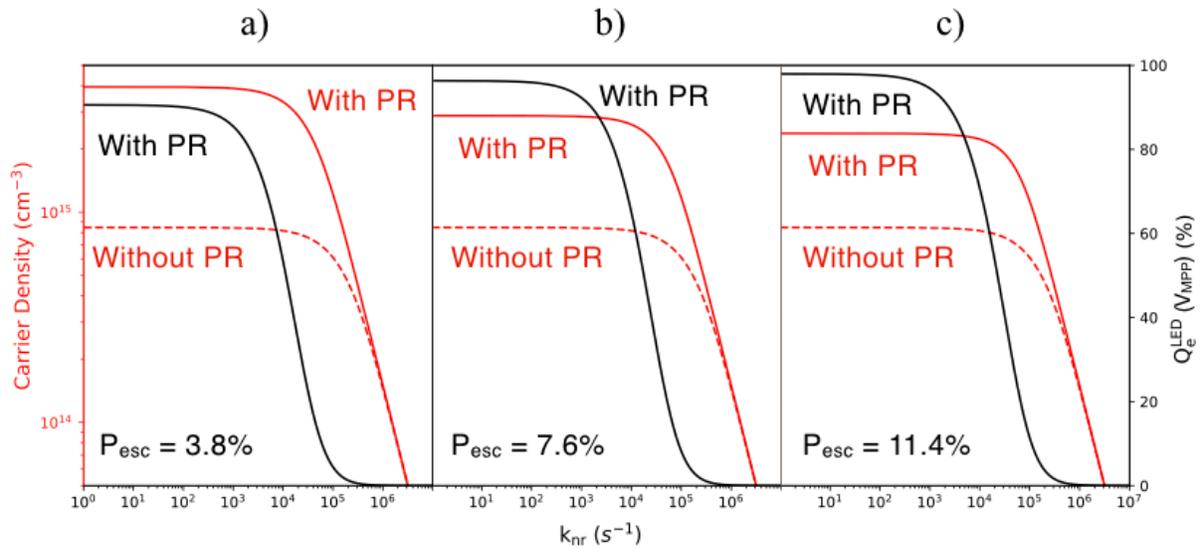

**Figure S21.** *The effect of PR on MPP steady-state carrier density and $Q_e^{\text{LED}}(V_{\text{MPP}})$ (calculated with an injection current achieved at a voltage bias of $V_{MPP}$) as a function of $k_1$ for a) $P_{esc}$ = 3.8%, b) 7.6%, and c) 11.4%.*